\documentclass[amsmath,amssymb,twocolumn,prd,showpacs,preprintnumbers]{revtex4}

\selectfont 

\usepackage{graphicx}       
\usepackage{dcolumn}        
\usepackage{bm}             
\usepackage{psfrag}
\usepackage{color}

\newcommand{\mass}{\mu}

\newcommand{\bU}{\mathbf{U}}
\newcommand{\bM}{\mathbf{M}}
\newcommand{\bR}{\mathbf{R}}
\newcommand{\balp}{\boldsymbol{\alpha}}
\newcommand{\bbet}{\boldsymbol{\beta}}
\newcommand{\bgam}{\boldsymbol{\gamma}}

\newcommand{\alp}{\alpha}

\newcommand{\gam}{\gamma}

\newcommand{\beq}{\begin{equation}}
\newcommand{\eeq}{\end{equation}}
\newcommand{\beqa}{\begin{eqnarray}}
\newcommand{\eeqa}{\end{eqnarray}}

\newcommand{\Op}{\hat{\mathcal{D}}_2}

\newcommand{\nn}{\nonumber}

\newcommand{\afn}{u}
\newcommand{\Pol}{\mathcal{P}}

\begin{document}


\title{Massive vector fields on the Schwarzschild spacetime: quasinormal modes and bound states}

\author{Jo\~ao G. Rosa}
 \email{joao.rosa@ed.ac.uk}
 \affiliation{%
 SUPA, School of Physics and Astronomy, University of Edinburgh, Edinburgh, EH9 3JZ, UK
}%

\author{Sam R. Dolan}
 \email{s.dolan@soton.ac.uk}
 \affiliation{%
 School of Mathematics, University of Southampton, Highfield, Southampton SO17 1BJ, UK}%

\date{\today}

\begin{abstract}
We study the propagation of a massive vector or Proca field on the Schwarzschild spacetime. The field equations are reduced to a one-dimensional wave equation for the odd-parity part of the field and two coupled equations for the even-parity part of the field. We use numerical techniques based on solving (scalar or matrix-valued) three-term recurrence relations to compute the spectra of both quasinormal modes and quasi-bound states, which have no massless analogue, complemented in the latter case by a forward-integration method. We study the radial equations analytically in both the near-horizon and far-field regions and use a matching procedure to compute the associated spectra in the small-mass limit. Finally, we comment on extending our results to the Kerr geometry and its phenomenological relevance for hidden photons arising e.g.~in string theory compactifications.
\end{abstract}

\maketitle


\section{Introduction}

The propagation of massless fields and linearized metric perturbations on black hole spacetimes has been extensively studied over the last half-century, following the foundational work of Regge and Wheeler \cite{Regge-Wheeler:1957}. Key motivations for such studies range from questions of black hole stability \cite{Vishveshwara:1970}, no-hair theorems \cite{Bekenstein:1972}, Hawking radiation \cite{Page:1976}, scattering and absorption \cite{Matzner:1968, Dolan:2008, Crispino:Dolan:Oliveira:2009}, and gravitational-wave dynamics \cite{Zerilli:1970}. In the 1970s, it was established that the response of a black hole to a generic initial perturbation is characterized by a spectrum of damped resonances \cite{Chandrasekhar:Detweiler:1975}, called quasinormal modes (for recent reviews see e.g.~\cite{Berti-Cardoso-Starinets, Konoplya:2011}) and that, after a period of quasinormal ringing, the perturbing field subsequently undergoes power-law decay \cite{Price:1972}. 

The propagation of massive fields, such as the electron, neutrino and baryonic fields, on black hole spacetimes has also received some attention \cite{Deruelle-1974, Damour-1976, Chandrasekhar:1976, Gal'tsov-1983, Galtsov, Gaina-1992, Simone-1992, Gaina-1993, Lasenby-2005-bs, Doran-2005, Dolan-2006, Laptev-2006, Grain-Barrau, KerrFermi, Koyama:2001, Koyama:2002, Jing:2005, KZ:2006, K:2006, KZM:2007}. If the field has mass $\mu$ and the black hole has mass $M$, the key dimensionless quantity is the mass coupling $M \mu \equiv G M \mu / \hbar c = M  \mu / m_P^2$ where $m_P$ is the Planck mass (henceforth we set $G = c = \hbar = 1$). For astrophysical black holes and known elementary fields we expect $M \mu$ to be large. For example, for a (hypothetical) neutrino mass $\mu_{\nu} \sim 0.02$ eV, one obtains $M \mu = 1.5 \times 10^8$ for solar-mass black holes. We therefore expect massive fields to behave very differently from massless fields in most astrophysical scenarios, at least for multipoles $l \lesssim M \mu$. 

Time domain studies \cite{Koyama:2001, Koyama:2002, Jing:2005, KZM:2007} have also revealed interesting differences in the propagation of massive and massless fields. First, quasinormal ringing is generally less easily excited, and the frequencies and decay times depend on $M \mu$. Second, power-law decay is replaced with a slowly-decaying, oscillating phase (reminiscent of Zitterbewegung) in which the frequency of oscillation is given by the mass of the field, rather than the mass of the black hole. It has been shown that, at late times $\mu t \gg 1 /  (M \mu)^2$, all multipoles exhibit a `universal' oscillatory decay, with a spin-independent envelope $\sim t^{-5/6}$ \cite{Koyama:2001, Koyama:2002, Jing:2005, KZM:2007} and frequency $\mu$. Third, massive fields with spin exhibit additional longitudinal radiative degrees of freedom, and low multipoles $l < s$ of massive fields (for example the vector field monopole) have a radiative character, rather than the static character apparent in the massless case. It has long been known that a black hole cannot support massive `hair', and consequently does not have a well-defined baryon number \cite{Bekenstein:1972}.

From the point of view of phenomenonology, the most interesting scenarios arise when $M\mu\sim 1$, a condition that may be attained either for light primordial black holes postulated to be produced during inflation \cite{Hawking, Zeldovich, Carr} or `ultralight' exotic particles found in beyond the Standard Model scenarios. These scenarios are particularly significant in the context of the well-established superradiant instability of rotating black holes \cite{Press-1972, Teukolsky:1973, Teukolsky:1974, Deruelle-1974, Damour-1976, Zouros:1979, Detweiler-1980, Cardoso:2004, Furuhashi:2004jk, Cardoso:2005vk, Dolan-2007, Konoplya:2008, Rosa, Cardoso:2011xi} subject to massive bosonic bound states. For the massive scalar field, numerical studies have found \cite{Furuhashi:2004jk, Cardoso:2005vk, Dolan-2007, Rosa} that the maximum growth rate of the instability is approximately $\tau^{-1} \sim 1.5 \times 10^{-7} (GM/c^3)^{-1}$ for a rapidly-rotating black hole at $a \sim 0.99 M$, and it occurs for a coupling $M \mu \sim 0.42$. Hence, this process is relevant in astrophysical environments for particles with masses $\mu\lesssim10^{-10}$ eV. Ultra-light particles have been proposed to be ubiquitous in string theory compactifications \cite{Arvanitaki-JMR}, where the several axionlike fields acquire masses exclusively through nonperturbative string instanton effects, making them exponentially sensitive to the size of the associated compact cycles. This leads to a generic landscape of ultralight axions, known as the `string axiverse' \cite{Arvanitaki-JMR, Arvanitaki, Kodama:Yoshino:2011}, populating all mass scales possibly down to the present Hubble scale, $H_0\simeq 10^{-33}$ eV. Recently, there have been suggestions that ultralight bosonic fields (e.g.~$\mu \sim 10^{-22}$ eV) may play the role of dark matter in galactic halos \cite{Lora-2011}. It is possible that rotating black holes may provide a unique probe of the ultralow energy spectrum of string theory compactifications, with an extremely rich phenomenology associated with the formation of bosonic superradiant bound states around astrophysical black holes, such as the emission of gravitational waves, gaps in the mass-spin black hole Regge spectrum and `bosenova-like' particle bursts \cite{Arvanitaki}.

Massive hidden $U(1)$ vector fields are also a generic feature of Beyond-Standard-Model scenarios and, in particular, string theory compactifications, arising in the latter case from a variety of sources, such as broken non-Abelian orbifolds in heterotic compactifications, and D-brane configurations and bulk Ramond-Ramond fields in type II string theories \cite{Goodsell, Jaeckel, Camara}. The nature of $U(1)$ vector masses is, however, inherently different from the axion case, where the associated shift symmetry (reminiscent of an underlying $U(1)$ symmetry) is only broken nonperturbatively. $U(1)$ gauge symmetries may, on the other hand, be broken by perturbative effects following the Higgs mechanism or by Stueckelberg-type couplings. In the former case, the resulting vector masses are generically determined by the soft-supersymmetry breaking terms in the hidden sector and hence are quite sensitive to the mechanism mediating supersymmetry breaking to this particular sector, which may yield a broad range of possible values. Light $U(1)$ particles can nevertheless be obtained in this case for hyperweak hidden sector gauge couplings, which naturally arise e.g.~from volume suppression if the hidden sector lives on D-branes wrapping a large compact cycle. Alternatively, one may envisage scenarios where the string coupling is actually extremely small in the overall compactification (within the limits of the `gravity as the weakest force' conjecture \cite{ArkaniHamed}), whereas the visible (Standard Model) sector wraps a collapsed cycle that makes the associated gauge couplings unnaturally large \footnote{We thank John March-Russell for pointing this out.}. 

Given such motivations, it is natural to speculate on the possibility that black hole physics, and in particular, the mechanism of unstable bosonic bound states driven by superradiance, can be used to probe the existence of ultralight hidden vector particles. For example, supermassive black holes, with masses $\sim 10^5 - 10^{10} M_{\text{solar}}$ are believed to be found in the centre of most active galaxies. Unstable `bosonic clouds' may form in the vicinity of rotating supermassive black holes, if particles with masses in the range $\sim10^{-16} - 10^{-21}$ eV exist. (Note that a reasonable upper bound on the mass of the photon is $m_\gamma<10^{-18}$ eV \cite{PDG}).  Stellar-mass black holes, formed in supernovae, are likely to be even more abundant; such black holes are sensitive to particles with masses $\sim 10^{-10} - 10^{-12}$ eV. Since the amplification mechanism relies on purely gravitational physics, it is largely independent of the kinetic mixing between the hidden and visible photons which is the basis for other types of analysis (see e.g.~\cite{Goodsell}). 

There are also avenues to explore within the context of Standard Model physics. For example, in astrophysical environments, the electromagnetic field can acquire a small `effective mass' via the Anderson-Higgs mechanism, induced by the accreting plasma surrounding the black hole \cite{Tamburini, Anderson}. This raises the possibility that superradiant instabilities may even play a role during black hole accretion.  As a step in this direction, the effect of strong magnetic fields on the instability timescale of the massive scalar field was considered in \cite{Konoplya:2008}. 

The study of massive vector field perturbations in rotating black hole spacetimes poses a challenging problem from both the analytical and numerical points of view, due to the apparent nonseparability of the Proca equations on the Kerr spacetime. Nonseparability is particularly frustrating because all other massless and massive fields (at least scalar and Dirac) admit separable solutions. 
Since it seems that any exploration of the phenomenology of the Proca field on the Kerr spacetime will require a careful analysis of {\em partial} (rather than ordinary) differential equations, it is essential that we first establish a solid understanding of the Proca field on the Schwarzschild spacetime, where the equations \emph{are} separable, which is the purpose of this work. Of course, since the Schwarzschild spacetime does not exhibit superradiance, we do not expect to find any instability in this case. 

Massive vector field perturbations on spherically-symmetric spacetimes have received some attention in the literature \cite{Galtsov, K:2006, KZM:2007, Herdeiro}, although comparatively less than other massive and massless perturbations. The problem of solving the massive vector field or Proca equations on the Schwarzschild spacetime is in itself quite challenging, given the coupling between the different components of the field that results from the broken gauge invariance and the additional longitudinal degree of freedom. Previous studies have examined the monopole mode \cite{K:2006, KZM:2007}, described by a single equation, and large multipoles \cite{Galtsov}, which can be studied using WKB methods. Very recently, the Proca field in higher-dimensional spherically symmetric spacetimes was examined in the context of Hawking radiation \cite{Herdeiro}. Here, we analyze Proca field perturbations on the Schwarzschild spacetime for generic multipoles, using both numerical and analytical techniques to focus on the spectrum of quasinormal (QN) modes and (quasi-)bound states. These spectra reveal several peculiar features that are related to both the massive and the higher-spin nature of the perturbations. To interpret our findings, we then compare our results with those for both the massive scalar and the (massless) electromagnetic fields.

This work is organized as follows. In Sec.~\ref{sec:analysis} we outline the properties of the Schwarzschild spacetime and formulate the Proca equations describing massive vector field perturbations on this geometry. In Sec.~\ref{subsec:separation} we separate the radial and angular parts of the spin-1 wavefunction using vector spherical harmonics, and we show that odd- and even-parity parts are completely decoupled. In Sec.~\ref{subsec:electromagnetic} we explore the relationship between the massless limit of the Proca field and the electromagnetic field. In Sec.~\ref{sec:QNBS} we define the massive spin-1 QN modes and bound states, and describe numerical methods for computing their spectra, based on continued-fraction and forward-integration techniques. We present a selection of numerical results for the spectra in Sec.~\ref{sec:numerical-results}. In Sec.~\ref{sec:approximations}, we develop analytical techniques to study the radial wavefunctions in the near- and far-field regions, and apply a functional matching procedure to obtain the associated spectra in the limit $M\mu\ll1$. We conclude with a summary of our main results in Sec.~\ref{sec:conclusion}, where we also discuss possible extensions and future work on this topic.


\section{Analysis\label{sec:analysis}}

\subsection{Spacetime and field equations\label{subsec:spacetime}}
The Schwarzchild spacetime is described by coordinates $x^{\mu} = \{ t, r, \theta, \phi \}$ and the line element
\begin{eqnarray}
ds^2=-f(r) dt^2 + f^{-1}dr^2 +r^2 \left( d\theta^2 + \sin^2 \theta d\phi^2 \right)~,
\end{eqnarray}
where $f(r) = 1 - 2M/r$. The exterior region ($r > r_H=2M$) is stationary and static; in other words there exists a timelike Killing vector ($\xi^\mu = \delta_0^\mu$) which is hypersurface-orthogonal. The spacetime is Ricci-flat, i.e.~$R=0$ and $R_{\mu \nu} \equiv {R^\gam}_{\mu \gam \nu} = 0$. 

The Proca field strength $F_{\mu \nu}$ is defined in terms of the (physical) vector potential $A_\nu$ by $F_{\mu \nu} = A_{\nu ; \mu} - A_{\mu ; \nu}$, where ${}_{;\mu}$ denotes a covariant derivative with respect to $x^\mu$. The field equations in vacuum are simply
\beq
{F^{\mu \nu}}_{;\mu} = \mass^2 A^\nu~.
\eeq
It follows that ${F^{\mu \nu}}_{;\mu \nu} = 0$ and hence the Lorenz condition $A^{\mu}_{;\mu} = 0$ is simply a consequence of the field equations. In other words, there is no gauge freedom in the Proca field, which thus describes three physical degrees of freedom. Hence, the field equations can be written as
\beq
{A^{\nu ; \mu}}_{ \mu} - \mass^2 A^\nu = 0~,  \label{field-eq}
\eeq
with the supplementary condition $A^\mu_{;\mu} = 0$. In the following sections, we will often take $M = 1$.

\subsection{Separation of variables\label{subsec:separation}}

To separate the angular part of the vector potential components, we introduce a basis of four vector spherical harmonics $Z_\mu^{(i)lm}$, defined as follows,
\begin{eqnarray} \label{vector-harmonics}
Z_{\mu}^{(1)lm} &=& \left[ 1, 0, 0, 0 \right] Y^{lm} \\
Z_{\mu}^{(2)lm} &=& \left[ 0, f^{-1}, 0, 0 \right] Y^{lm} \\
Z_{\mu}^{(3)lm} &=& {r\over\sqrt{l(l+1)}} \left[ 0, 0, \partial_\theta, \partial_\phi  \right] Y^{lm} \\
Z_{\mu}^{(4)lm} &=&  {r\over\sqrt{l(l+1)}} \left[ 0, 0, \frac{1}{s_\theta} \partial_\phi, - s_\theta \partial_\theta \right] Y^{lm}~,   
\end{eqnarray}
where $Y^{lm} \equiv Y^{lm}(\theta, \phi)$ denote the ordinary (scalar) spherical harmonics and $s_\theta\equiv\sin\theta$. Note that the four vector spherical harmonics are defined in an analogous way to the ten tensor spherical harmonics used for describing gravitational perturbations in the Lorenz gauge \cite{Barack:Lousto:2005}. These harmonics satisfy the orthogonality conditions
\beq
\int \left(Z_{\mu}^{(i)lm}\right)^\ast \eta^{\mu \nu} Z_\nu^{(i')l'm'} d\Omega = \delta_{ii'} \delta_{ll'} \delta_{mm'}~,
\eeq
where we defined $\eta^{\mu\nu}=\text{diag}[1, f^2, 1/r^2, 1/(r^2\sin^2\theta)]$ and $d\Omega=\sin\theta d\theta d\phi$. 

Let us briefly consider the transformation properties of these harmonics under parity inversion, $\mathbf{x} \rightarrow -{\mathbf x}$, i.e. $\theta \rightarrow \pi - \theta$ and $\phi \rightarrow \phi + \pi$, under which the spatial components of a vector field transform as $[A_r, A_\theta, A_\phi ] \rightarrow [-A_r, +A_\theta, -A_\phi ]$. The first three harmonics ($i = 1, 2, 3$) in Eq.~(\ref{vector-harmonics}) then pick up a sign of $(-1)^{l}$ under parity inversion, whereas the last ($i=4$) harmonic picks up the opposite sign $(-1)^{l+1}$. It is standard to call the former the `even-parity' or `electric' modes and the latter the `odd-parity' or `magnetic' modes.

We may now decompose the vector potential in this basis:
\beq
A_{\mu}(t,r,\theta,\phi) = \frac{1}{r} \sum_{i=1}^{4} \sum_{lm} c_i \, \afn^{lm}_{(i)}(t,r) Z_\mu^{(i)lm}(\theta, \phi)~,
\label{sep-ansatz}
\eeq
where $c_1 = c_2 = 1$, $c_3 = c_4 = [l(l+1)]^{-1/2}$. This {\it ansatz} is sufficient to separate the vector field equations (\ref{field-eq}) into a set of four second-order partial differential equations in $r$ and $t$,
\begin{eqnarray}
\Op \afn_{(1)} &+& \left[ \frac{2}{r^2} \left(  \dot{\afn}_{(2)} - \afn^\prime_{(1)} \right) \right] = 0 \label{eq-alp1} \\
\Op \afn_{(2)} &+& \frac{2}{r^2}\left[ \left(  \dot{\afn}_{(1)} - \afn^\prime_{(2)} \right) - f^2\left( \afn_{(2)} - \afn_{(3)} \right) \right]=0 \nonumber\\
&~&  \label{eq-alp2}\\
\Op \afn_{(3)} &+& \left[ \frac{2 f l (l+1)}{r^2} \afn_{(2)} \right] = 0 \label{eq-alp3}  \\
\Op \afn_{(4)} &=& 0~.   \label{odd-parity}
\end{eqnarray}
Here $\dot\afn \equiv \tfrac{\partial \afn}{\partial t}$, $\afn^\prime \equiv \tfrac{\partial \afn}{\partial r^\ast}$ and the tortoise coordinate $r_\ast$ is defined via $dr_\ast = f^{-1} dr$. The differential operator $\Op$ is given by
\beq
\Op \equiv -\frac{\partial^2}{\partial t^2} + \frac{\partial^2}{\partial r_\ast^2} - f \left[ \frac{l(l+1)}{r^2} + \mu^2 \right]~.
\eeq
The fourth equation (\ref{odd-parity}), which describes the odd-parity sector, is completely decoupled from the first three equations  (\ref{eq-alp1})--(\ref{eq-alp3}), which describe the even-parity sector. The even-parity equations must be supplemented by the Lorenz condition,
\beq
-\dot{\afn}_{(1)} + \afn^\prime_{(2)} + \frac{f}{r} \left(\afn_{(2)} - \afn_{(3)}  \right)  =  0~.  \label{Lorenz-eq}
\eeq
This condition may be used to reduce the even-parity system to a pair of coupled differential equations. We may then replace Eq.~(\ref{eq-alp2}) with
\begin{eqnarray}
\Op \afn_{(2)} - \frac{2 f}{r^2} \left(1 - \frac{3}{r} \right) \left( \afn_{(2)} - \afn_{(3)} \right) = 0   \label{eq-alp2-alt}
\end{eqnarray}
and note that Eqs.~(\ref{eq-alp3}) and (\ref{eq-alp2-alt}) form a closed system. Hence, we have obtained one decoupled wave equation for the odd-parity part of the vector potential and two coupled wave equations for the even-parity modes.


\subsection{Proca field and the electromagnetic limit\label{subsec:electromagnetic}}
Let us consider the relation between the Maxwell field and the massless limit of the Proca field.

\subsubsection{The monopole mode}
In the special case of the monopole mode ($l=0$, even-parity), for which only the first two ($i=1$ and $2$) harmonics are defined, we obtain a single decoupled equation
\beq
\left[ -\frac{\partial^2}{\partial t^2} + \frac{\partial^2}{\partial r_\ast^2}  - f \left( \frac{2(r-3)}{r^3} + \mass^2 \right) \right] \afn_{(2)} = 0~.  \label{monopole-eq}
\eeq
This is precisely the monopole equation investigated by Konoplya (see Eq.~(15) in \cite{K:2006}). The Lorenz condition (\ref{Lorenz-eq}) implies that $\dot \afn_{(1)} =  \tfrac{f}{r} \partial_r (r \afn_{(2)})$.

In the massless limit ($\mass \rightarrow 0$) we may apply a `Chandrasekhar transform' \cite{Chandrasekhar:1975, Chandrasekhar:1983} to the monopole equation to show that $\dot{\afn}_{(1)}$ satisfies a scalar wave equation, 
\beq
\left[-\frac{\partial^2}{\partial t^2}  +  \frac{\partial^2}{\partial r_\ast^2}  - \frac{2 f}{r^3} \right] \dot \afn_{(1)} = 0~.
\eeq
In the electromagnetic limit, this degree of freedom is not physical and can be removed by a gauge transformation. In other words, in electromagnetism the monopole part of the field does not have radiative degrees of freedom and, in the Lorenz gauge, $A_{\mu}$ can be written as the sum of a gauge mode and a static field, i.e.~
\beq
A_{\mu} = \chi_{, \mu} + \frac{q}{r} \delta_{\mu}^0~,   \label{A-monopole-static}
\eeq
where $\chi$ is an arbitrary scalar field satisfying $\Box \chi = 0$. 

In the Proca case ($\mass \neq 0$), the static mode in Eq.~(\ref{A-monopole-static}) is \emph{not} a solution of the equation, whereas the radiative degree of freedom \emph{is} physical. In the small-mass limit, we then expect the spectrum of the monopole part of the Proca field, governed by (\ref{monopole-eq}), to approach the spectrum of the $l=0$ mode of a scalar field.


\subsubsection{Massless limit of odd-parity modes: Price's equation}
In the massless limit, the odd-parity equation (\ref{odd-parity}) reduces to Price's equation \cite{Price:1972},
\beq
\Op (r^2 \phi_1) = 0~,
\eeq
which governs the evolution of the Maxwell scalar of spin-weight zero $\phi_1$. In fact, inserting the {\it ansatz} in Eq.~(\ref{sep-ansatz}) for $A_\mu$ into the definition of $\phi_1 \equiv F_{\mu \nu} (l^{\mu} n^{\nu} + \bar{m}^\mu m^{\nu})$, where $l^\mu$, $n^\mu$, $m^{\mu}$ and $\bar{m}^{\mu}$ are the null vectors of the Kinnersley tetrad \cite{Kinnersley}, leads to
\beq
\phi_1^{lm} = i \, \frac{l (l+1)}{r^2} \afn_{(4)}^{lm}(t,r) Y_{lm}(\theta, \phi)~.
\eeq


\subsubsection{Massless limit of even-parity $l > 0$ modes}

The pair of coupled wave equations for the even-parity sector, Eqs.~(\ref{eq-alp3}) and (\ref{eq-alp2-alt}), can be combined into a single fourth-order equation. In the massless case, $\mass = 0$, it is notable that the fourth-order equation for $\afn_{(3)}$ can be factorized as follows:
\beq
\frac{1}{rf^3} \hat{\mathcal{D}}_{RW}^{l,s=1} \left[ f^{-1} \hat{\mathcal{D}}_{RW}^{l,s=0} \left( r \afn_{(3)} \right) \right] = 0~, \label{factorization}
\eeq
where we have defined the generalized `Regge-Wheeler' operator $\hat{\mathcal{D}}_{RW}^{l,s}$ as
\beq
\hat{\mathcal{D}}_{RW}^{l,s} \equiv - \frac{\partial^2}{\partial t^2} + \frac{\partial^2}{\partial r_\ast^2} - f \left( \frac{l(l+1)}{r^2} + \frac{2 (1-s^2)}{r^3} \right)~.
\eeq
The equation for $\afn_{(2)}$ can also be factorized, although the operators do not have such a simple form. 

It can be shown that if $r \afn_{(3)}$ satisfies the `inner' scalar wave equation in Eq.~(\ref{factorization}), i.e.~$\hat{\mathcal{D}}_{RW}^{l,s=0} (r \afn_3) = 0$, then the corresponding vector potential is pure-gauge, i.e.
\beq
A_\mu = \chi_{,\mu} , \quad \quad \text{where} \quad \chi = \frac{\afn_{(3)}}{l (l+1)} Y_{lm}(\theta, \phi)~.
\eeq
In other words, in the electromagnetic case, the `inner' degree of freedom has no physical significance since it can be completely removed by a gauge transformation $A_{\mu} \rightarrow A_{\mu} - \chi_{, \mu}$. However,for $\mu \neq 0$ there is no gauge freedom and so this degree of freedom becomes physical. We expect its spectrum to approach that of a scalar wave in the low-mass limit, as for the monopole mode. 

The `outer' wave equation in Eq.~(\ref{factorization}) for the quantity $\psi \equiv f^{-1} \hat{\mathcal{D}}_{RW}^{l,s=0} \left( r \afn_{(3)} \right) $ has physical significance even in the massless case. It is straightforward to show that $\psi$ can be written as
\beq
\psi = \afn^\prime_{(3)} - \frac{l(l+1)}{r} \afn_{(2)}~.  \label{psi-def}
\eeq
Working in reverse,  inserting Eq.~(\ref{psi-def}) into Eqs.~(\ref{eq-alp3}) and (\ref{eq-alp2-alt}) leads to a `vector' ($s=1$) wave equation for $\psi$,
\beq
\hat{\mathcal{D}}_{RW}^{l,s=1} \psi = \Op \psi  = 0~.  \label{psi-eq}
\eeq
We can conclude that, in the electromagnetic limit, the physically-meaningful even-parity and odd-parity degrees of freedom are described by the same dynamical equation [see Eqs.~(\ref{odd-parity}) and (\ref{psi-eq})] and hence have the same spectrum. However, this degeneracy is broken if the field has mass; in addition, the gauge degrees of freedom (in the monopole and in higher modes) acquire physical significance in the Proca case. In the following sections, we investigate the richer structure of the Proca spectrum.


\section{Frequency spectra: Quasinormal Modes and Bound States\label{sec:QNBS}}

In this section, we investigate the spectrum of characteristic modes of the Proca equation on the Schwarzschild spacetime. We start with a frequency-domain representation, where
\beq
\afn_{(i)}^{lm}(t,r) = u_{(i)}^{lm} (\omega, r) e^{- i \omega t}
\eeq
and $\omega$ may take complex values. Note that $\text{Im}(\omega) < 0$ corresponds to exponential decay, while $\text{Im}(\omega) > 0$ leads to exponential growth, such that we expect to find only the former behaviour in the Schwarzschild spacetime. 

Physical modes on a black hole spacetime must be purely ingoing at the horizon from the point of view of a local observer. The ingoing condition corresponds to the horizon boundary condition
\beq \label{ingoing}
u_{(i)}( \omega, r) \sim e^{-i \omega r_\ast}~, 
\eeq
as $r_\ast \rightarrow - \infty$. In the asymptotically flat region, the solution resembles 
\beq \label{asymptotic}
u_{(i)}(\omega, r)\sim B_{(i)}(\omega) e^{-k r_\ast} + C_{(i)}(\omega) e^{+ k r_\ast}~, 
\eeq
as $r_\ast \rightarrow +\infty$, where $B_{(i)}(\omega)$, $C_{(i)}(\omega)$ are complex coefficients and we define $k = \sqrt{\mu^2 - \omega^2}$ such that $\text{Re}(k) > 0$. There are two types of special mode that we may consider: (i) the QN modes, defined by $B_{(i)}(\omega) = 0$, which asymptotically resemble purely outgoing waves; (ii) the quasibound states, defined by $C_{(i)}(\omega) = 0$, which are spatially localized within the vicinity of the black hole, i.e.~decay exponentially away from the black hole. In either case, imposing an asymptotic boundary condition generates a discrete spectrum of allowed frequencies, $\{ \omega_{ln}^{(\Lambda)}(M \mass) \}$. Here, each QN mode or bound-state frequency is labeled by its angular momentum number $l \ge 0 $, overtone number $n \ge 0$, and `polarization' numbers $\Lambda$, which we further describe below. 


\subsection{Continued-fraction method\label{subsec:ctd-frac}}

As described in \cite{Dolan-2007}, either frequency spectrum may be found using a method which relies on the fact that, with a suitable {\it ansatz}, the desired solutions of the differential equations correspond to minimal solutions of three-term recurrence relations for series coefficients, which may be found by solving a continued-fraction equation \cite{Leaver-1985}. An appropriate {\it ansatz} is
\beq \label{qnbs-ansatz}
\afn_{(i)}(\omega,r) = f^{-2i\omega} r^{-\nu} e^{qr} \sum_n a^{(i)}_n [f(r)]^n~,   
\eeq
where $\nu = (\omega^2-q^2)/q$. To seek QN frequencies, we set $q = k$ [i.e.~$q = \sqrt{\mu^2 - \omega^2}$ and thus $\text{Re}(q) > 0$]. On the other hand, for bound-state frequencies we set $q = -k$ [i.e.~$\text{Re}(q) < 0$]. Inserting Eq.~(\ref{qnbs-ansatz}) into the governing equations (\ref{eq-alp3}), (\ref{odd-parity}), (\ref{eq-alp2-alt}) and (\ref{monopole-eq}) then leads to three-term relations of the form
\begin{eqnarray}
\alpha_0 a_1 + \beta_0 a_0 &=& 0 \\
\alpha_n a_{n+1} + \beta_n a_n + \gamma_n a_{n-1} &=& 0, \quad n > 0~.
\end{eqnarray}
Where the equations are decoupled, i.e.~for the odd-parity and monopole cases, $\alpha_n, \beta_n, \gamma_n$ are scalar quantities, and the QN or bound-state frequencies are those for which
\beqa
&&\beta_n - \frac{ \alpha_{n-1} \gamma_n}{\beta_{n-1} - \frac{\alpha_{n-2}\gamma_{n-1}}{\beta_{n-2} - \alpha_{n-3} \gamma_{n-2} / \ldots} } =   \nonumber\\
&=&\frac{\alpha_n \gamma_{n+1}}{\beta_{n+1} - \frac{\alpha_{n+1} \gamma_{n+2}}{\beta_{n+2} - \alpha_{n+2} \gamma_{n+3} / \ldots }}~.
\eeqa
Methods for solving continued-fractions have been described in detail elsewhere \cite{Leaver-1985} and this method is extensively used (see e.g.~Sec.~IIIH in \cite{Konoplya:2011} or \cite{Zhidenko-thesis}), so that we will give no further details here. Where the equations are not decoupled, i.e.~for even-parity modes, $\alpha_n$ etc.~are matrix-valued. This case is discussed in Sec.~\ref{subsec:ctdfrac-even}. 


\subsubsection{Monopole mode}
The monopole mode $l=0$ is governed by Eq.~(\ref{monopole-eq}). The QN mode spectrum was found by Konoplya \cite{K:2006} via the continued-fraction method. We obtain the following coefficients:
\beqa
\alpha_n &=& (n + 1)(n + 1 - 4 i \omega)~, \\
\beta_n   &=& -2 n^2 - 2 q^{-1} \left( 4 i \omega q + 3 q^2 - \omega^2 + q \right) n  \nn \\
 && + q^{-1} \big(4 i \omega^3 + 12 q \omega^2 - 12 i \omega q^2 - 4 q^3 + \nn \\
 && + q - \omega^2 + 4 i \omega q + 3 q^2  \big)~, \\
\gamma_n &=& n^2 - q^{-1} \left( 4i \omega q + 2 q^2 - 2 \omega^2 \right) n +\nn \\
&& +q^{-2} \left(\omega - iq\right)^4 - 4~.
\eeqa
%

\subsubsection{Odd-parity modes}

The odd-parity modes satisfy the decoupled wave equation (\ref{odd-parity}) which reduces to the $s=1$ Regge-Wheeler (RW) equation $\hat{\mathcal{D}}_{RW}^{l,s=1} \afn_{(i)} = 0$ in the massless limit. In this case, we obtain $\alpha_n$ as above and
\begin{eqnarray}
\beta_n &=& - 2 n^2 + 2 q^{-1} \left( 4 i \omega q + 3 q^2 - \omega^2 - q \right) n  \nn \\
 &&  + q^{-1} \big[ 3 q^2 + (12 q - 1) \omega^2 - 4 q^3 - q l (l+1) -\nn\\
&& - 12 i \omega q^2 + 4 i \omega q + 4 i \omega^3 \big] ,  \\
\gamma_n &=&  n^2 - 2 q^{-1} \left( q^2 - \omega^2 + 2i \omega q \right) n -\nn \\
&-&\!\!q^{-2}\! \big[\!-4i\omega q^3\!+\!4 i \omega^3 q\!+\!6 \omega^2 q^2\!-\! q^4\!-\!\omega^4\! +\!q^2\big]~.
\end{eqnarray}
%


\subsubsection{Even-parity modes\label{subsec:ctdfrac-even}}

The even-parity modes satisfy a pair of coupled differential equations, Eqs.~(\ref{eq-alp3}) and (\ref{eq-alp2-alt}). Inserting the {\it ansatz} (\ref{qnbs-ansatz}) into these equations then leads to a \emph{matrix-valued} three-term recurrence relation,
\begin{eqnarray}
\balp_0 \bU_{1} + \bbet_0 \bU_{0} &=& 0~, \\
\balp_n \bU_{n+1} + \bbet_n \bU_{n} + \bgam_n \bU_{n-1} &=& 0, \quad n > 0~, 
\end{eqnarray}
with a vectorial coefficient $\mathbf{U}_n = \begin{pmatrix} a^{(2)}_n \\ a^{(3)}_n \end{pmatrix}$ and matrices:
\begin{eqnarray}
\balp_n &=& \begin{pmatrix} \alp_n & 0 \\ 0 & \alp_n \end{pmatrix}, \quad
\bbet_n = \begin{pmatrix} \beta_n + 1 & -1 \\ 2 l (l+1) & \beta_n \end{pmatrix}~, \nn \\
\bgam_n &=& \begin{pmatrix} \gamma_n - 3 & 3 \\ 0 & \gamma_n  \end{pmatrix}, 
\end{eqnarray} 
where
\beqa
\alp_n &=& (n + 1)(n + 1 - 4i\omega)~, \nn \\
\beta_n &=& -2n^2 + \left[ \frac{6q^2 + 8iq\omega - 2\omega^2}{q} - 2  \right] n  \nn \\
&-& l(l+1) +{1\over q}\big[ -12 i \omega q^2 + 3 q^2- 4 q^3\nn \\
&-&  \omega^2 + 12 q \omega^2 + 4 i \omega q + 4 i \omega^2 \big]~, \\
\gamma_n &=& \left(n + \frac{(\omega-iq)^2}{q} - 1 \right)\left( n + \frac{(\omega-iq)^2}{q} + 1 \right)~.\nn
\eeqa

The matrix-valued three-term recurrence relation can be solved using matrix-valued continued fractions \cite{SWP:1999}. We search for roots of the equation $\bM \bU_0 = 0$, where
\beq
\bM \equiv   \bbet_0  -  \balp_0 \left[ \bbet_1 - \balp_1\left( \bbet_2 + \balp_2 \bR_2^{+} \right) \bgam_2  \right]^{-1} \bgam_1~,
\eeq
with $\bU_{n+1}=\bR_n^+\bU_n$ and
\beq
\bR_n^+ = - \left( \bbet_{n+1} + \balp_{n+1} \bR_{n+1}^+ \right)^{-1} \bgam_{n+1}~.
\eeq
For nontrivial solutions $\bU_0$, one must hence solve
\beq
\text{det} \left| \bM \right| = 0~. \label{eq:det}
\eeq
In nondegenerate cases, we expect to find two independent solutions, with distinct eigenvectors $\bU_0^{(a)}$ and $\bU_0^{(b)}$. To distinguish between the solutions, we define the quantity
\beq
\Pol \equiv \lim_{r_\ast \rightarrow \infty} \frac{ \afn_{(3)}(\omega, r)} {\afn_{(2)}(\omega, r)} =  \sum_{n=0}^\infty a^{(3)}_n / \sum_{n=0}^\infty a^{(2)}_n~,    \label{eq-pol}
\eeq
which we will loosely refer to as the polarization.


\subsection{Forward-integration method\label{subsec:forward-integration}}

An alternative numerical method can be used to compute the spectrum of bound states, based on the expected convergence of the solutions at infinity. In the vicinity of the horizon, the expansion in Eq.~(\ref{qnbs-ansatz}) for the radial functions satisfying ingoing boundary conditions may be written as
\beq \label{horizon_expansion}
u_{(i)}(\omega,r)=(r-r_H)^{-2i\omega}\sum_{n=0}^\infty b_{(i)n}(r-r_H)^n~.
\eeq
The series coefficients $b_{(i)n}$ for $n\geq1$ can be easily found, as a function of the leading coefficient $b_{(i)0}$, by substituting this {\it ansatz} into the radial equations~(\ref{eq-alp3}), (\ref{odd-parity}) and (\ref{eq-alp2-alt}) and using a symbolic algebra package, e.g.~Mathematica. We may then use Eq.~(\ref{horizon_expansion}) as an initial condition close to the horizon to numerically integrate the radial equations up to the far-field region.

For the odd-parity modes, we expect this to lead to an asymptotic solution of the form Eq.~(\ref{asymptotic}), i.e. a linear combination of an exponentially divergent and an exponentially convergent term, with the former vanishing for bound states. We may thus determine the bound-state spectrum by setting $b_{(4)0}=1$ and minimizing the resulting $u_{(4)}(\omega,r)$ for an arbitrarily large distance $r\gg r_H$ in the complex $\omega$-plane. 

This method can be extended for the even-parity modes, where the mode equations are coupled. In this case, we may obtain a family of numerical solutions at infinity parametrized by the unknown leading coefficients $(b_{(2)0}, b_{(3)0})$, with bound states corresponding to particular values of the latter that lead to pure exponentially decaying solutions. The associated spectrum may then be computed by choosing a suitable basis for the space of initial coefficients, for example $(b_{(2)0}, b_{(3)0})=(1,0)$ and $(0,1)$, and defining a $2\times2$ matrix of solutions  
\beq \label{asymptotic_matrix}
\mathbf{S}(\omega,r)  =
\begin{pmatrix}
u_{(2)}^{(1,0)}(\omega,r) & u_{(2)}^{(0,1)}(\omega,r)   \\ 
u_{(3)}^{(1,0)}(\omega,r) & u_{(3)}^{(0,1)}(\omega,r)  
\end{pmatrix}~.
\eeq
The particular linear combinations of the near-horizon solutions corresponding to bound states will thus correspond to the kernel of $\mathbf{S}$ evaluated at $r\rightarrow \infty$. In practice, this corresponds to minimizing $\det|\mathbf{S}|$ in the complex $\omega$-plane at an arbitrarily large distance. Notice that, for each minimum, only the eigenvector associated with an asymptotically vanishing eigenvalue will correspond to a physical solution, whereas the remaining eigenstate is unphysical and yields an exponentially large eigenvalue. This allows one to reconstruct the radial wave functions in each case and determine the associated polarization, as defined in Eq.~(\ref{eq-pol}).

Although this method cannot be applied to compute the QN mode spectrum, where the purely divergent nature of the solutions is hard to determine numerically, it can be more easily implemented e.g.~with Mathematica, where the lengthy algebraic expressions resulting from the multiple matrix inversions required by the continued-fraction method are rather difficult to minimize. Furthermore, forward-integration provides an independent check of the numerical results obtained with the latter method, making our analysis more robust.


\section{Numerical results\label{sec:numerical-results}}

\subsection{Quasinormal modes}

\begin{figure}[htbp]
\includegraphics[scale=0.85]{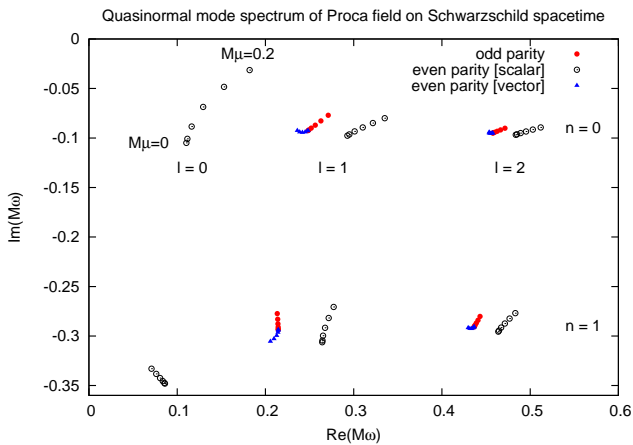}
\caption{Quasinormal mode frequencies of the Proca field, for $l=0$ (monopole), $l=1$ (dipole) modes, $l = 2$ (quadrupole) modes, for a range of field masses $M \mu = 0, 0.04, \ldots, 0.2$. The fundamental ($n=0$) and first overtones ($n=1$) are shown. For a given $l$, $n$ there are two even-parity modes, and one odd-parity mode. In the massless limit, the `scalar' even-parity mode has the same QN frequency as the scalar ($s=0$) field, whereas the `vector' even-parity and odd-parity modes have the same QN frequency as the electromagnetic field.}
\label{fig:qnfreq}
\end{figure}

Fig.~\ref{fig:qnfreq} shows the effect of mass on the quasinormal frequencies of the low-$l$, $n$ modes of the Proca field. For a given angular momentum number $l$ and overtone number $n$, there are two even-parity modes and one odd-parity mode. The even-parity modes may be distinguished by their behaviour in the massless limit, as discussed in Sec.~\ref{subsec:electromagnetic}. In this limit, the spectrum of `scalar' modes (which are unphysical pure-gauge modes in electromagnetism) reduces to the spectrum of a scalar field.  In the same limit, the `vector' even-parity and odd-parity modes are degenerate, with the frequencies of the electromagnetic field. The field mass breaks this degeneracy.

The plot in Fig.~\ref{fig:qnfreq} illustrates the effect of the field mass upon the QN spectrum of low-$l$, low-$n$ modes. As expected, this is greatest for the lowest modes. As previously observed by Konoplya \cite{K:2006}, the decay rate of the monopole mode decreases substantially as the mass coupling $M \mu$ is increased. However, the physical relevance of the QN mode also diminishes as the mass increases, as the height of the effective potential barrier decreases.

\begin{figure}[htbp]
\includegraphics[scale=0.68]{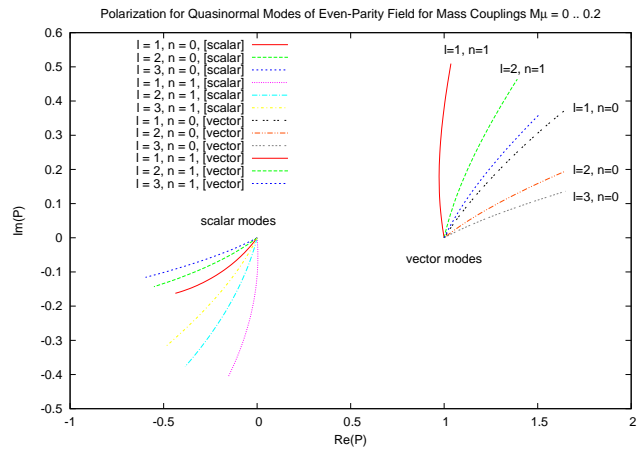}
\caption{Polarization state of even-parity quasinormal modes. The plot shows the complex number $\Pol$, i.e.~the ratio $\afn_{(3)} / \afn_{(2)}$ far from the black hole [defined in Eq.~(\ref{eq-pol})], as a function of the mass coupling $M\mu = 0 \ldots 0.2$, for $l=1,2,3$ and $n = 0,1$. For scalar (vector) modes, $\Pol \rightarrow 0$ ($\Pol \rightarrow 1$) as $M\mu \rightarrow 0$.}
\label{fig:pol-qn}
\end{figure}

In the Proca case ($\mu \neq 0$), the two even-parity QN modes  (`scalar' and `vector') have distinct frequencies and polarization states. In Fig.~\ref{fig:pol-qn}, we examine the polarization of the modes at large distances, by plotting $\Pol$ as defined in Eq.~(\ref{eq-pol}) as a function of $M\mu$, for $l=1,2,3$ and for the fundamental mode ($n = 0$) and the first overtone ($n = 1$). In the massless limit, $\Pol$ approaches $0$ for `scalar' modes and $1$ for `vector' modes. Additionally, one observes that $\Pol$ varies smoothly with $M \mu$. In theory, a measurement of $\Pol$ and $\omega_{ln}$ would allow one to independently deduce the mass of the black hole $M$ and the mass of the field $\mass$. In practice, however, detecting QN ringing from a Proca field is unlikely to be possible in the foreseeable future.

\subsection{Bound states}

As discussed in \cite{Deruelle-1974, Damour-1976, Detweiler-1980, Gal'tsov-1983, Gaina-1992, Gaina-1993, Lasenby-2005-bs, Laptev-2006, Dolan-2007, KerrFermi}, a massive field may be localized in the vicinity of a black hole in (quasi-)bound states with complex frequencies.  In this section we present a selection of numerical results for the Proca-field bound-state spectrum. We have verified that the results obtained via the continued-fraction method (Sec.~\ref{subsec:ctd-frac}) are in excellent agreement with those obtained with the forward-integration technique (Sec.~\ref{subsec:forward-integration}). 

The bound states of the Proca field were previously considered in Ref.~\cite{Galtsov} for large multipoles, where it was shown that in the limit $M\mass \rightarrow 0$, the spectrum is hydrogenic, i.e.~\footnote{We correct the expression obtained in \cite{Galtsov} by introducing a factor of 2 that was missing in Eq.~(6) of this work.}
\beq
\text{Re} \left( \omega / \mu \right) \approx 1 - \frac{(M\mass)^2}{2 N^2}~, \label{reE}
\eeq
where $N = j + 1 + n$ and $j=l+S$ is the total angular momentum of the state as measured by an asymptotic observer, with spin projection $S=0, \pm1$. Our results are fully consistent with this if $S=+1$ for the monopole mode, in agreement with the rules for addition of angular momenta, such that $|l-1|\leq j\leq l+1$ for spin-1 fields. 

\begin{figure}[htbp]
\includegraphics[scale=0.65]{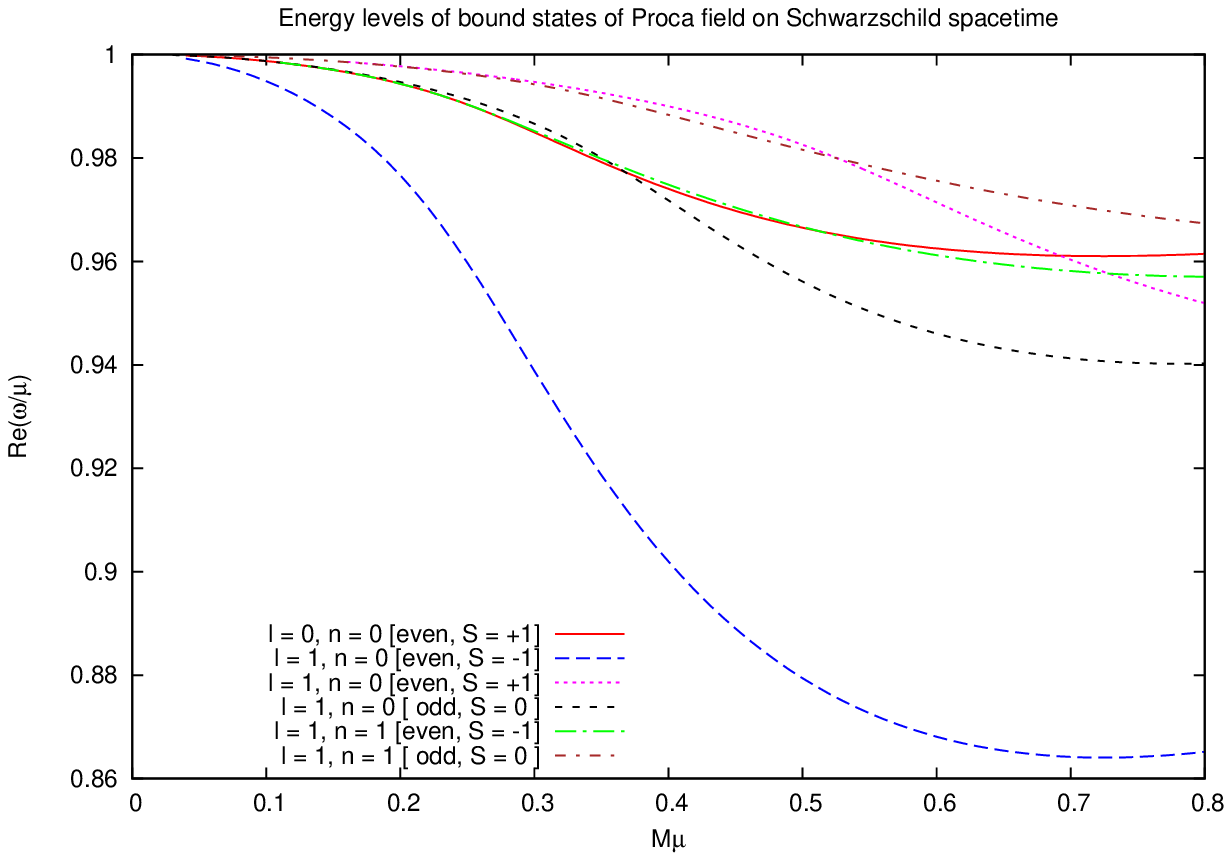}
\includegraphics[scale=0.65]{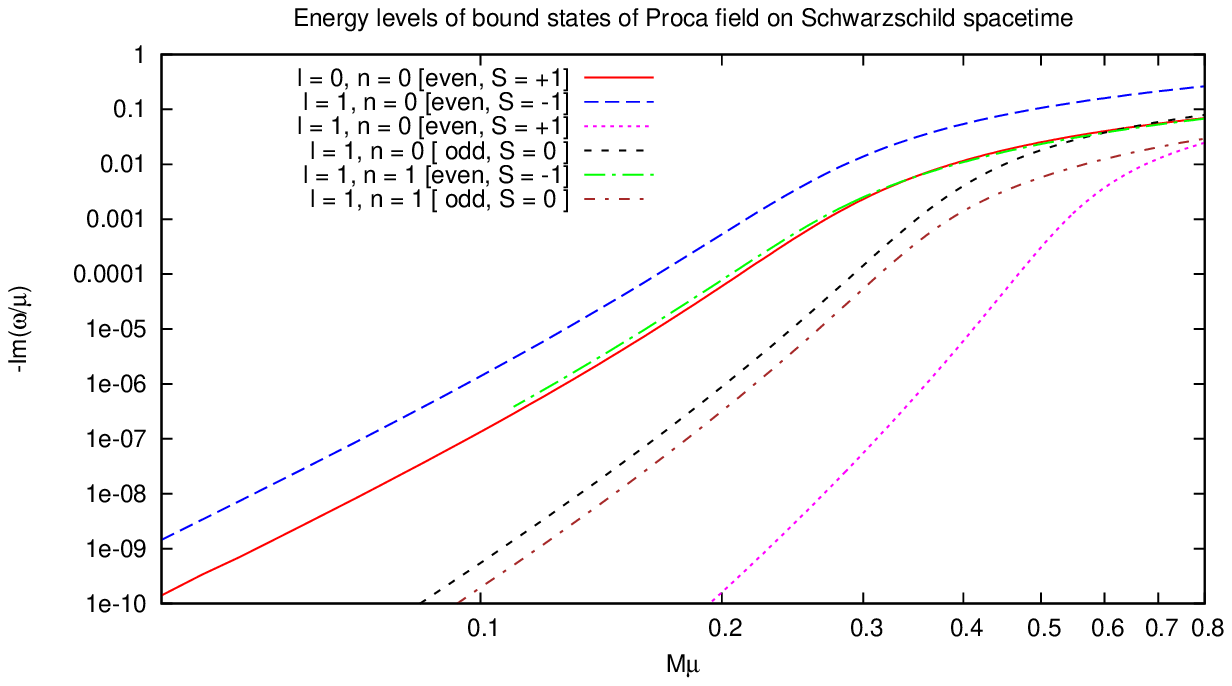}
\caption{Bound state levels of the Proca field on the Schwarzschild spacetime. The upper plot shows the real part of the frequency $\text{Re}(\omega / \mu)$ as a function of the mass coupling $M \mu$, and the lower plot shows (the negative of) the imaginary part $\text{Im}(\omega / \mu)$ on a logarithmic scale. The modes are labeled by their angular momentum number $l$, overtone number $n$, spin projection $S$ and parity (odd or even).}
\label{fig:energy_levels}
\end{figure}

In Fig.~\ref{fig:energy_levels} we show the bound-state frequency spectrum $\omega / \mu$ as a function of the mass coupling $M \mu$, for the lowest modes $l=0$, $1$. For a given $l$ and $n$, there are three types of mode: (i) odd parity, $S = 0$, (ii) even parity, $S = +1$ and (iii) even parity, $S = -1$, the monopole being a type (ii) mode. As predicted by Eq.~(\ref{reE}), the lowest-energy mode is $l=1$, $S = -1$.  

\begin{figure}[htbp]
\includegraphics[scale=0.974]{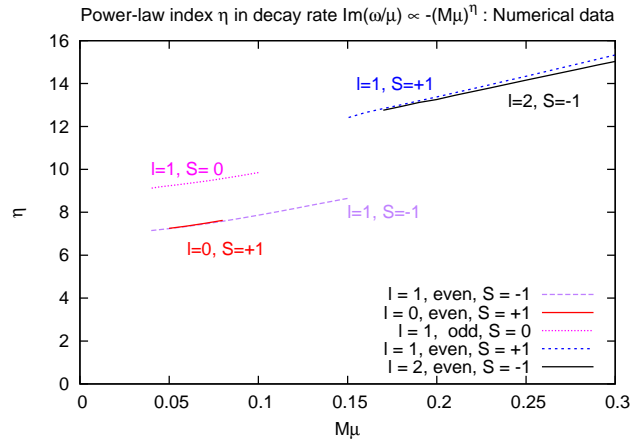}
\caption{Numerical data for the exponent $\eta$ in the power-law relationship $\text{Im} (\omega / \mu) \propto -(M\mu)^{\eta}$ which determines the decay rate of the quasi-bound states in the small-coupling regime $M\mu \ll l$. Here $l = 0,1, \ldots$ is the angular momentum number, and $S \in \{-1, 0, +1\}$ is the spin projection of the state in the large-$r$ regime. The data shown were obtained by numerically evaluating the function $\eta =  -\frac{\partial \ln[\text{Im}(\omega / \mu)]}{\partial \ln [ M \mu ]}$. Note that numerical evaluation becomes increasingly difficult in the small-$M\mu$ regime, where $|\text{Im}(\omega / \mu)|$ is tiny ($\lesssim 10^{-12}$). The data strongly suggests that modes $l=L, S=+1$ and $l=L+1, S=-1$ share the same exponent $\eta$. The data is consistent with Eq.~(\ref{eq-exponent}), which implies that, in the limit $M \mu \rightarrow 0$, the exponent tends to $\eta = 7$ ($l=0,\ S=+1$ and $l=1,\ S=-1$), $\eta = 9$ ($l=1,\ S=0$) and $\eta = 11$ ($l=1,\ S=+1$ and $l=2,\ S=-1$). }
\label{fig:imag-exponent}
\end{figure}

The imaginary part of the frequency, which sets the decay rate of the mode, increases monotonically with $M \mu$. The lower plot of Fig.~\ref{fig:energy_levels} shows that, in the small-$M\mu$ regime, there is a power-law dependence, $\text{Im}(\omega / \mass) \propto -(M\mass)^\eta$, where $\eta$ depends on $l$ and spin projection $S$, and the constant of proportionality depends on the overtone number. From the numerical data we infer that
\beq
\eta=4l+2S+5. \label{eq-exponent}
\eeq
For example, Fig.~\ref{fig:imag-exponent} shows the exponent $\eta$ estimated from the numerical data for $M \mu\ll1$, clearly showing that the modes $l=L, S=+1$ and $l=L+1, S=-1$ have the same exponent $\eta$. Our data is \emph{not} in agreement with the results found in \cite{Galtsov}, suggesting that the latter analysis is not applicable to the lowest multipoles. 

\begin{figure}[htbp]
\includegraphics[scale=0.79]{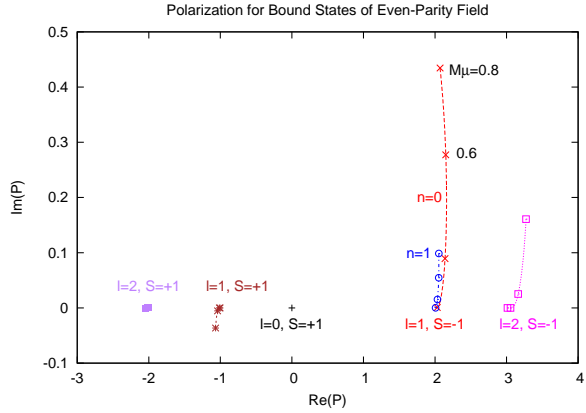}
\caption{Polarization state of even-parity bound states. The plot shows the complex number $\Pol$, i.e.~the ratio $\afn_{(3)} / \afn_{(2)}$ evaluated asymptotically [see Eq.~(\ref{eq-pol})], as a function of the mass coupling $M\mu = 0 \ldots 0.8$, for a selection of the lowest modes ($l=0,1,2$) and overtones ($n = 0,1$). The points show the values of $\Pol$ at $M \mu = 0, 0.2, 0.4, 0.6$ and $0.8$. In the limit of vanishing mass, we find $P \rightarrow -l$ for $S = +1$ modes, and $\Pol \rightarrow l+1$ for $S = -1$ modes (where $S$ is the spin projection).}
\label{fig:pol_bs}
\end{figure}

The pair of even-parity modes may be distinguished by examining the polarization $\Pol$ defined in Eq.~(\ref{eq-pol}). For $S = +1$ modes we find $P \rightarrow -l$ as $M\mass \rightarrow 0$, and for $S = -1$ modes we find $\Pol \rightarrow l + 1$ in the same limit, as shown in Fig.~\ref{fig:pol_bs}.

\begin{figure}[htbp]
\includegraphics[scale=0.852]{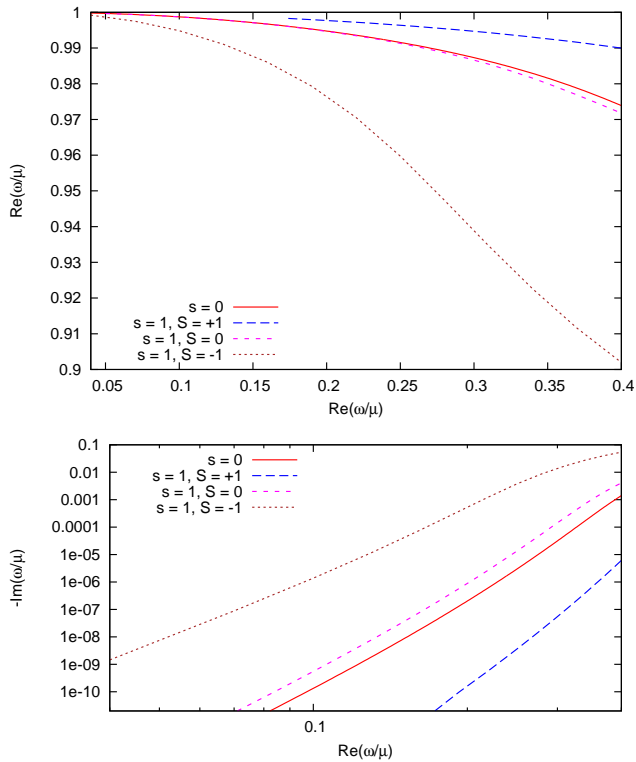}
\caption{Comparison between the bound-state frequencies of the $l=1,~n=0$ modes for scalar (spin-0) and Proca (spin-1) fields. The imaginary part of the frequency exhibits the same dependence on the mass coupling $M\mu$ for both the scalar and longitudinal vector (odd-parity) bound states, being larger for the latter, while for the transverse vector (even-parity) states it grows with distinct powers of $M\mu$, being larger (smaller) for $S=-1$ ($S=+1$) [see Eq.~(\ref{eq-exponent})].}
\label{fig:bs-s01-l1}
\end{figure}

Fig.~\ref{fig:bs-s01-l1} shows the comparison between the spectrum of odd-parity Proca modes (with $S=0$) and the spectrum of the massive scalar-field ($s=0$) modes. In both cases, the exponent is $\eta = 4l + 5$, although the constant of proportionality is larger for Proca-field perturbations.

\begin{figure}[htbp]
\includegraphics[scale=0.63]{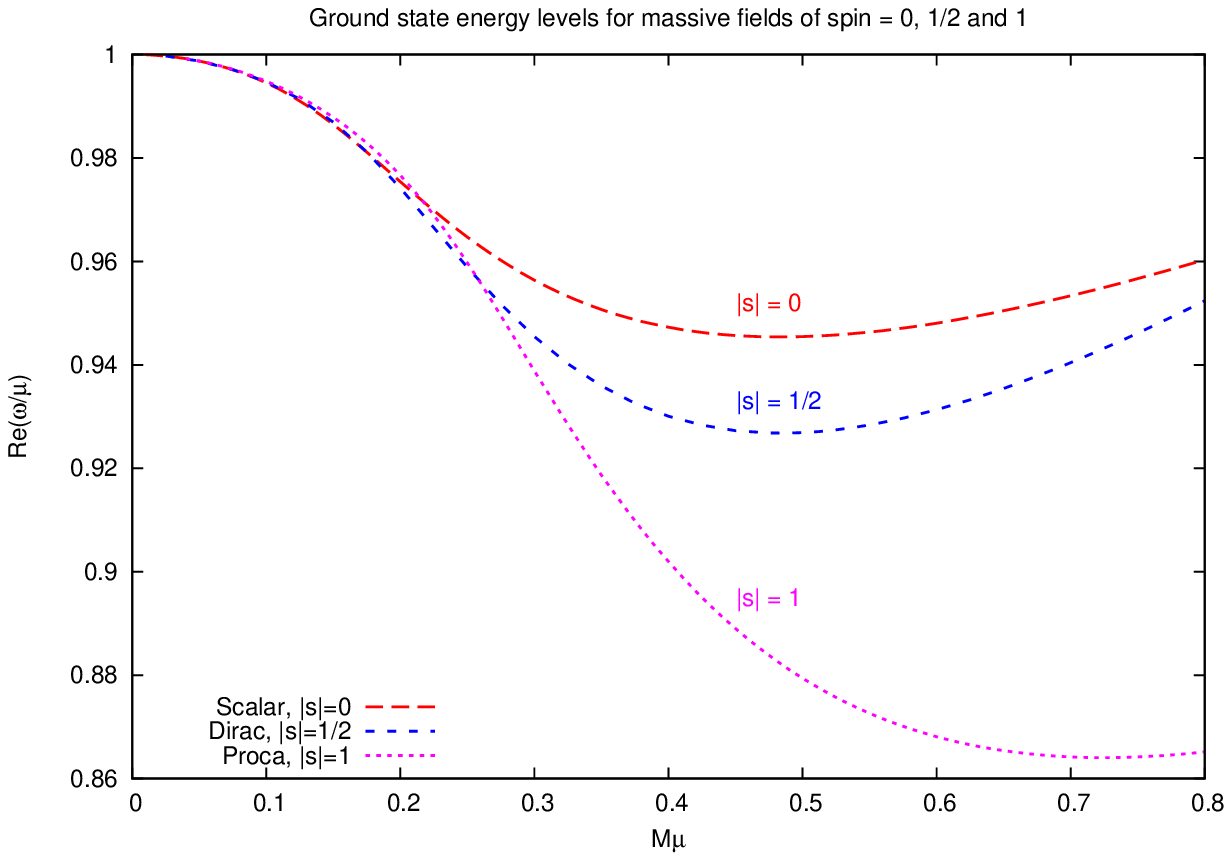}
\includegraphics[scale=0.63]{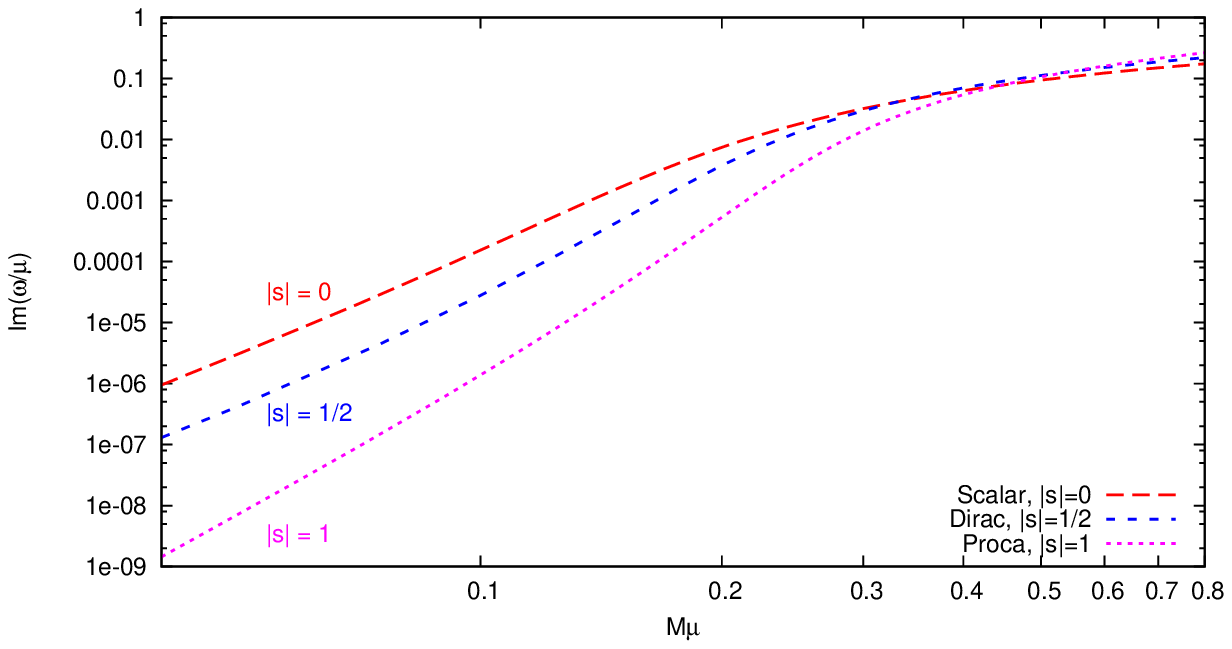}
\caption{Ground state energy levels of massive fields of spin $s=0$ (scalar), $s=1/2$ (Dirac), and $s = 1$ (Proca) on the Schwarzschild spacetime. The upper plot shows the real part of frequency $\text{Re}(\omega / \mu)$ as a function of the mass coupling $M \mu$ and the lower plot shows (the absolute value of) the imaginary part of frequency $\text{Im}(\omega / \mu)$, which sets the decay rate, on a logarithmic scale.}
\label{fig:energy_levels_spins}
\end{figure}

Figure \ref{fig:energy_levels_spins} compares the ground state frequencies of the Proca field ($l=1$, $S = -1$) with those of the massive Dirac \cite{Lasenby-2005-bs} and scalar \cite{Dolan-2007} fields. In the small-coupling limit, we see that, regardless of spin, all fields exhibit a hydrogenic spectrum, $\omega/\mu \approx 1- (M\mu)^2 / 2$. For low couplings, $M\mu \lesssim 0.4$, the Proca field is more stable than the other fields, i.e.,~decays more slowly. At larger couplings, we see that the Proca field exhibits the largest binding energy ($\text{Re}(\omega/\mu - 1) \approx -0.14$), although here the lifetime of the state is actually only a few black-hole light-crossing times.

\section{Analytical results for small-mass coupling\label{sec:approximations}}

To better understand the numerical results obtained in the previous section, we now study the equations for the odd- and even-parity modes in the limit of small mass $\mu$ and small frequency $\omega$. The basic approach is to derive separate solutions in the near-horizon and far-field regions, defined by $\mu x, \omega x\ll l$ and $x\gg1$, respectively, where  $x=r/2M-1$, and then match them in their common domain of validity, $1\ll x\ll l/\omega$. As this requires $\omega \ll l$, we expect this to give a better approximation to the numerical results for bound states, for which $\omega\sim\mu\ll1$, than for the QN modes, where $\omega\gtrsim \mathcal{O}(0.1)$ even for $\mu\rightarrow0$. This approach will nevertheless give us a better insight into the analytical form of the solutions and the massless limit, being complementary to the more precise numerical analysis performed earlier.

We start by writing the equations for the perturbations in terms of the dimensionless variable $x$, yielding for the odd-parity modes:
\begin{eqnarray} \label{mode_equations_odd}
&&\big[x^2(x+1)^2\partial_x^2+x(x+1)\partial_x+V(x)\big]u_{(4)}=0~,\nn \\
\end{eqnarray}
where $V(x)=4\omega^2(x+1)^4-4\mu^2x(x+1)^3-\lambda^2x(x+1)$, $\lambda^2=l(l+1)$ and both $\omega$ and $\mu$ are given in units of the inverse black hole mass $M^{-1}$. For the even-parity modes, one finds the coupled second-order differential equations:
\begin{eqnarray} \label{mode_equations_even}
&&\big[x^2(x+1)^2\partial_x^2+x(x+1)\partial_x+V(x)\big]u_{(2)}=\nn \\
&&\qquad\qquad\qquad\qquad=x(2x-1)\big(u_{(2)}-u_{(3)}\big)~, \nonumber\\
&&\big[x^2(x+1)^2\partial_x^2+x(x+1)\partial_x+V(x)\big]u_{(3)}=\nn \\
&&\qquad\qquad\qquad\qquad=-2\lambda^2x(x+1)u_{(2)}~,
\end{eqnarray}

As we will see below, it will be convenient to write this system in terms of $\psi(x)$, defined in Eq.~(\ref{psi-def}), which up to a constant rescaling may be written as
\begin{equation} \label{psi}
 \psi={x\partial_x u_{(3)}-\lambda^2u_{(2)}\over x+1}~.
\end{equation}
Replacing $u_{(2)}$ by this function, one obtains
\begin{eqnarray} \label{mode_equations_even_psi}
&&\big[x^2(x+1)^2\partial_x^2+x(x+1)\partial_x+V(x)\big]\psi= \nn \\
&&\qquad\qquad\qquad\qquad=4\mu^2x(x+1)u_{(3)}~,\nn\\
&&\big[x^2(x+1)^2\partial_x^2\!+\!x(x+1)(2x+1)\partial_x\!+\!V(x)\big]u_{(3)}= \nn \\
&&\qquad\qquad\qquad\qquad=2x(x+1)^2\psi~.
\end{eqnarray}
This explicitly shows that the equation for $\psi$ decouples in the massless case, giving the physical `vector' mode solutions described earlier, whereas for $\psi=0$ one obtains a decoupled equation for $u_{(3)}$ that yields the unphysical `scalar' modes. Let us now analyze the behaviour of the odd- and even-parity solutions in more detail.


\subsection{Odd-parity modes}

\begin{figure}[htbp]
\includegraphics[scale=0.90]{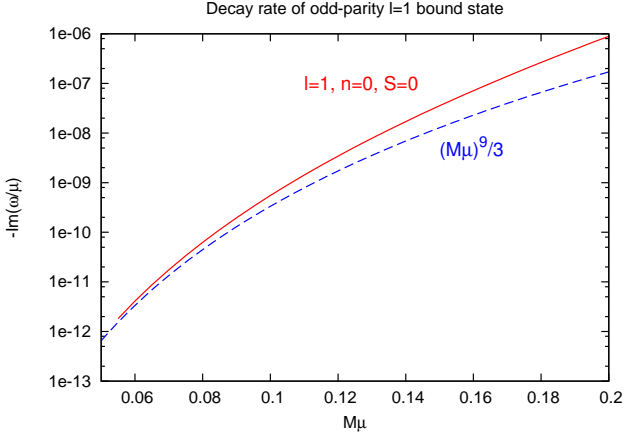}
\caption{Comparison between the numerical and analytical results for the imaginary part of the bound-state frequency for the odd-parity $l=1,n=0$ mode as a function of the mass coupling $M\mu$. The solid [red] line shows numerical data, and the dashed [blue] line shows the analytical approximation $\text{Im}(\omega / \mu) \approx - \tfrac13 (M\mu)^9$.}
\label{fig:odd_bound}
\end{figure}

In the near-region, Eq.~(\ref{mode_equations_odd}) reduces to
\begin{eqnarray} \label{mode_equations_odd_near}
\big[x^2(x+1)^2\partial_x^2+x(x+1)\partial_x+4\omega^2
-\lambda^2x(x+1)\big]u_{(4)}&&\nn \\
=0~,~&&
\end{eqnarray}
which has a general solution given in terms of hypergeometric functions
\begin{eqnarray} \label{odd_solution_near}
u_{(4)}^{near}&=&A_{(4)}x^{-2i\omega}(x+1)^{1+\delta}\times \nn \\
&&\!\!\!\!\!\!\!\!{}_2F_1(-l-2i\omega+\delta,l+1-2i\omega+\delta,1-4i\omega,-x)+\nn \\
&&\!\!\!\!\!\!\!\!+B_{(4)}x^{2i\omega}(x+1)^{1+\delta}\times \nn \\
&&\!\!\!\!\!\!\!\!{}_2F_1(-l+2i\omega+\delta,l+1+2i\omega+\delta,1+4i\omega,-x)~,\nn \\
\end{eqnarray}
where $\delta=\sqrt{1-4\omega^2}$. It can be easily seen in terms of the tortoise coordinate $r_*$ that ingoing solutions at the horizon require setting $B_{(4)}=0$, and using the asymptotic properties of the hypergeometric function \cite{Abramowitz} we can derive the $x\gg1$ form of the near-region solution
\begin{eqnarray} \label{odd_solution_near_large}
u_{(4)}^{near}&\simeq& A_{(4)}\Gamma[1-4i\omega]\times\nn \\
&&\bigg[{\Gamma[2l+1]\over\Gamma[l+1-2i\omega+\delta]\Gamma[l+1-2i\omega-\delta]}x^{l+1}+\nn \\
&&{\Gamma[-2l-1]\over\Gamma[-l-2i\omega+\delta]\Gamma[-l-2i\omega-\delta]}x^{-l}\bigg]~.
\end{eqnarray}
In the far-field region, Eq.~(\ref{mode_equations_odd}) can be written as
\begin{eqnarray} \label{mode_equations_odd_far}
[x^2\partial_x^2-4q^2x^2+4q\nu x-\lambda^2]u_{(4)}=0~,
\end{eqnarray}
where $q$ and $\nu$ were defined earlier [see Eq.~(\ref{qnbs-ansatz})], and the general solution can be written in terms of confluent hypergeometric functions
\begin{eqnarray} \label{odd_solution_far}
u_{(4)}^{far}&=&e^{-z/2}\big[C_{(4)}z^{l+1}M(l+1-\nu,2l+2,z)+\nn \\
&+&D_{(4)}z^{-l}M(-l-\nu,-2l,z)\big]~,
\end{eqnarray}
where $z=4qx$. For bound states, the linear combination which is regular at infinity corresponds to
\begin{eqnarray} \label{odd_solution_far_bound}
u_{(4)}^{bound}=\tilde{C}_{(4)}e^{-z/2}z^{l+1}U(l+1-\nu,2l+2,z)~,
\end{eqnarray}
which for $z\ll1$ takes the form \cite{Abramowitz}
\begin{eqnarray} \label{odd_solution_far_bound_small}
u_{(4)}^{bound}&\simeq&\tilde{C}_{(4)}{(4q)^{l+1}\pi\over\sin{(2l+2)\pi}}\bigg[{x^{l+1}\over\Gamma[-l-\nu]\Gamma[2l+2]}-\nn \\
&-&{(4q)^{-2l-1}\over\Gamma[l+1-\nu]\Gamma[-2l]}x^{-l}\bigg]~.
\end{eqnarray}
Thus, the near- and far-region solutions yield the same power-law behaviour in the intermediate region, and one can equate the associated coefficients to get the matching condition
\begin{eqnarray} \label{odd_bound_matching}
{\Gamma[-l-\nu]\Gamma[2l+2]\over\Gamma[l+1-\nu]\Gamma[-2l]}&=&-(4q)^{2l+1}{\Gamma[-2l-1]\over\Gamma[2l+1]}\times\nn \\
&&\!\!\!\!\!\!\!\!\!\!\!\!\!\!\!\!\!\!\!\!\!\!\!\!\!\!\!\!\!\!\!\!\!\!\!\!\!
\times{\Gamma[l+1-2i\omega+\delta]\over\Gamma[-l-2i\omega+\delta]}{\Gamma[l+1-2i\omega-\delta]\over\Gamma[-l-2i\omega-\delta]}~.
\end{eqnarray}
This condition can be solved in a similar way to \cite{Furuhashi:2004jk}, taking into account that the left hand side vanishes to leading order for $q\ll1$, corresponding to the poles of $\Gamma[l+1-\nu]$. These are given by $\nu=l+1+n$ for a non-negative integer $n$, yielding to lowest order in $M\mu$ a spectrum of Hydrogen-like bound states, as in the scalar-field case, and in agreement with our numerical results
\begin{eqnarray} \label{odd_bound_spectrum}
\omega\simeq\mu\bigg(1-{\mu^2\over 2(l+1+n)^2}\bigg)~.
\end{eqnarray}
We may then expand both the left and right-hand sides of the matching condition about this value to get the next-to-leading order (NLO) correction to the spectrum. This is not as straightforward as for scalar-field perturbations, given that there is an uncancelled pole in one of the $\Gamma$ functions. This can be overcome by taking to lowest order $\Gamma[l+1-2i\omega+\delta]\simeq \Gamma[l+2]$ and $\Gamma[-l-2i\omega+\delta]\simeq \Gamma[-l+1]$, which is in fact consistent with the approximations used in the near-region. We may proceed as in \cite{Furuhashi:2004jk} to obtain the next-to-leading order correction, valid for $l\geq1$
\begin{eqnarray} \label{odd_bound_imag}
\delta\omega&\simeq&{4^{2l+1}\mu^{4l+5}\over (l+1+n)^{2l+4}}{(2l+1+n)!\over n!}{(l+1)!(l-1)!\over [(2l)!(2l+1)!]^2}\nn \\
&\times&(1+2i\omega)\prod_{k=1}^l(k^2-1-4i\omega)~,
\end{eqnarray}
where $\omega$ takes the leading order value in Eq.~(\ref{odd_bound_spectrum}). From this one can extract the imaginary part of the bound-state frequency, which to leading order grows as $\mu^{4l+6}$ as for scalar-field perturbations, although with a different coefficient. For example, for the lowest-lying dipole mode, $l=1, n=0$, we obtain $\omega_I\sim\mu^{10}/3$, which is twice the value obtained for scalar-field perturbations \cite{Furuhashi:2004jk, {Detweiler-1980}}. In this sense, we classify the odd-parity modes as longitudinal bound states, which behave like scalar-field perturbations far from the black hole but have a vectorlike near-horizon behaviour, which makes the lowest-lying modes decay more rapidly. We illustrate this behaviour in Fig.~\ref{fig:odd_bound}, where one can see that the numerical curve approaches the matching result for small masses. 


For the QN modes, the coefficients $C_{(4)}$ and $D_{(4)}$ in Eq.~(\ref{odd_solution_far}) can be obtained by taking into account that $M(a,b,z)\simeq1$ for small $z$ and then matching them with the corresponding coefficients in Eq.~(\ref{odd_solution_near_large}). Asymptotically, the far-region solution takes the form
\begin{eqnarray} \label{odd_solution_far_QNM}
u_{(4)}^{QN}&\simeq& e^{-z/2}(-1)^{-\nu+l+1}z^\nu\times\nn\\
&\times&\bigg[C_{(4)}{\Gamma[2l+2]\over\Gamma[l+1+\nu]}-D_{(4)}{\Gamma[-2l]\over\Gamma[-l+\nu]}\bigg]+\nonumber\\
&+&e^{z/2}z^{-\nu}\bigg[C_{(4)}{\Gamma[2l+2]\over\Gamma[l+1-\nu]}+D_{(4)}{\Gamma[-2l]\over\Gamma[-l-\nu]}\bigg]~.\nn \\
\end{eqnarray}
Recalling that $z=4qx=4i\sqrt{\omega^2-\mu^2}$, it is easy to see that the first term corresponds to ingoing waves from infinity and should be set to zero. One should notice that for QN modes one expects $\nu\ll l$ in the limit $\mu\ll\omega\ll l$, as opposed to the case of bound states where $\nu$ takes integer values as seen above. We may thus consistently neglect $\nu$ and obtain the QN mode condition
\begin{eqnarray} \label{odd_QNM_matching}
1&=&(4q)^{2l+1}{\Gamma[-2l]\over\Gamma[-l]}{\Gamma[l+1]\over\Gamma[2l+2]}{\Gamma[-2l-1]\over\Gamma[2l+1]}\nn\\
&\times&{\Gamma[l+1-2i\omega+\delta]\over\Gamma[-l-2i\omega+\delta]}{\Gamma[l+1-2i\omega-\delta]\over\Gamma[-l-2i\omega-\delta]}.
\end{eqnarray}
We may then proceed as for bound states to approximately eliminate the uncancelled $\Gamma$ pole, obtaining to leading order, for $l\geq1$
\begin{eqnarray} \label{odd_QNM_matching_2}
&&1=-i(4\sqrt{\omega^2-\mu^2})^{2l+1}(l+1)!(l-1)!\nn \\
&&\times\left({l!\over (2l)!(2l+1)!}\right)^2
(1+2i\omega)\prod_{k=1}^l(k^2-1-4i\omega)~.\nn \\
\end{eqnarray}
The solutions of this equation in the lower half of the complex plane then yield the spectrum $\omega_{(l,n)}^{odd}$ of QN modes for odd-parity perturbations. For example, for $l=1$, solving this to leading order in $\omega$ and $\mu$ gives
\begin{eqnarray} \label{odd_QNM}
\omega_{(1,0)}^{odd}\simeq 0.515(1-i)+0.364(1+i)\mu^2~.
\end{eqnarray}
This result overestimates the real and imaginary part obtained numerically, which was expected given that the matching procedure only holds for $\omega\ll l$, but it nevertheless gives the correct qualitative $\mu$-dependence of the QN mode frequencies. Notice also that this does not agree with the matching analysis for massless vector fields performed in \cite{Panchapakesan:1987} using the Teukolsky equation rather than the spin-1 Regge-Wheeler equation. However, this analysis also departs significantly from the numerical data, which again is symptomatic of the failure of the approximations involved.


\subsection{Even-parity modes}

The problem becomes more involved for the even-parity modes, where one has a system of coupled differential equations and analytical results are harder to extract using the matching procedure described above. In the far-region, however, the system may be diagonalized and a general solution can be written as
\begin{eqnarray} \label{even_solutions_far_general}
u_{(2,3)}(z)=\sum_{S=\pm1}c_{(2,3)}^Su_{(S)}(z)~,
\end{eqnarray}
where $u_{(S)}$ satisfies the confluent hypergeometric equation:
\begin{eqnarray} \label{mode_equations_even_far}
[x^2\partial_x^2-4q^2x^2+4q\nu x-j(j+1)]u_{(S)}=0~,
\end{eqnarray}
with $j=l+S$ as defined earlier, and $c_{(3)}^+/c_{(2)}^+=-l$, $c_{(3)}^-/c_{(2)}^-=l+1$. The general solution is analogous to the odd-parity far-region solution, yielding:
\begin{eqnarray} \label{even_solution_far}
u_{(S)}^{far}&=&e^{-z/2}\big[C_{(S)}z^{j+1}M(j+1-\nu,2j+2,z)+\nn\\
&+&D_{(S)}z^{-j}M(-j-\nu,-2j,z)\big]~,
\end{eqnarray}
with bound states corresponding to the confluent hypergeometric function $U(j+1-\nu,2j+2,z)$.

In the near region, one could hope that the solutions would converge to the massless case expressions, given that the mass term can be neglected in this limit. However, writing the system of equations as in Eq.~(\ref{mode_equations_even_psi}), one explicitly sees that the effects of the vector field mass may only be neglected if the functions $\psi$ and $u_{(3)}$ are of comparable magnitude, which {\it a priori} is not necessarily the case. We may nevertheless investigate the form of these solutions neglecting the mass term and check whether a matching procedure can be used.

The generic solution in the near-region for $\psi$ is then identical to that obtained for $u_{(4)}$, yielding for ingoing boundary conditions at the horizon:
\begin{eqnarray} \label{even_solution_near_vector}
\psi^{near}&=&A_\psi x^{-2i\omega}(x+1)^{1+\delta}\times\nn \\
&&\!\!\!\!\!\!\!\!{}_2F_1(-l-2i\omega+\delta,l+1-2i\omega+\delta,1-4i\omega,-x)~,\nn \\
\end{eqnarray}
constituting the `vector' solution described earlier. One may also derive the form of the `scalar' solutions in the near-region by setting $\psi=0$ and solving the decoupled equation for $u_{(3)}$, which may also be written in terms of a hypergeometric function:
\begin{eqnarray} \label{even_solution_near_scalar}
u_{(3)}^{near}&=&A_{(3)} x^{-2i\omega}(x+1)^{2i\omega}\nn \\
&\times&{}_2F_1(-l,l+1,1-4i\omega,-x)~.
\end{eqnarray}
As expected, this corresponds to the solution for scalar-field perturbations in the near-region. For large $x$, these solutions behave like \cite{Abramowitz}:
\begin{eqnarray} \label{even_solution_near_large}
\psi^{near}&\simeq& A_\psi\Gamma[1-4i\omega]\times\nn \\
&&\!\!\!\!\!\!\bigg[{\Gamma[2l+1]\over\Gamma[l+1-2i\omega+\delta]\Gamma[l+1-2i\omega-\delta]}x^{l+1}+\nn\\
&+&{\Gamma[-2l-1]\over\Gamma[-l-2i\omega+\delta]\Gamma[-l-2i\omega-\delta]}x^{-l}\bigg]~,\nonumber\\
u_{(3)}^{near}&\simeq&A_{(3)}\Gamma[1-4i\omega]\times\nn \\
&&\!\!\!\!\!\!\bigg[{\Gamma[2l+1]\over\Gamma[l+1]\Gamma[l+1-4i\omega]}x^{l}+\nn \\
&+&{\Gamma[-2l-1]\over\Gamma[-l]\Gamma[-l-4i\omega]}x^{-l-1}\bigg]~.
\end{eqnarray}

One may then use the far-region solutions in Eq.~(\ref{even_solutions_far_general}) to compute these quantities in the far-region and take the limit $z\ll1$ as for the odd-parity modes. It is then easy to see that matching is only possible for $\nu\ll l$ which, from our analysis of the odd-parity modes, is expected to be the case for QN modes but not for bound states, where $\nu$ takes positive integer values.

That a matching between the (massless) near-region solutions and those in the far-field region is not possible for bound states is not completely unexpected, since these modes have no electromagnetic analogue as massless waves cannot be bound in a gravitational field. Furthermore, the form of the far-region solutions actually suggests a nontrivial mixing between the `vector' and `scalar' solutions of the massless case that cannot be determined using this approach.

Nevertheless, were the near-region solutions fully known, one could construct a matching condition analogous to Eq.~(\ref{odd_bound_matching}) for each of the far-region solutions $u_{(\pm)}$, labeled by $j=l\pm1$. Thus, one expects the bound-state spectrum to be given, to lowest order, by the poles of $\Gamma[j+1-\nu]$, which yield the hydrogenic spectrum in Eq.~(\ref{reE}), in agreement with the numerical results. Recall that in the limit $M\mu\ll1$ we had obtained $\Pol=-l,~l+1$, which is consistent with pure $u_{(\pm)}$ bound-state solutions far from the black-hole horizon. The polarization data suggests, however, that as $M\mu$ increases the mixing between these solutions becomes larger. 

The imaginary part of the even-parity bound-state modes depends, unfortunately, on the particular form of the near-region solution. One would naively expect it to grow like $\mu^{4j+6}=\mu^{4l+4S+6}$, in analogy with the odd-parity case, but as discussed in the previous section one can infer an additional factor of $\mu^{-2S}$ from the numerical data, which is clearly suggestive of a nontrivial mixing of the `vector' and `scalar' near-region solutions. 

On the other hand, for QN modes one may in principle take the limit $\nu\ll l$, where the confluent hypergeometric functions in Eq.~(\ref{even_solution_far}) can be written in terms of modified Bessel functions \cite{Abramowitz}
\begin{eqnarray} \label{modified_bessel}
M(n+1,2n+2,z)&=&\nn \\
&&\!\!\!\!\!\!\!\!\!\!\!\!\!\!\!\!\!\!\!\!\!\!\!\!\!\!\!\!\!\!\!\!\!\!\!\!\!\!\!\!\!\!\!
\Gamma[n+1/2]e^{z/2}\bigg({z\over4}\bigg)^{-n-1/2}I_{n+1/2}\bigg({z\over2}\bigg)
\end{eqnarray}
and we may write the general solution as:
\begin{eqnarray} \label{even_solution_far_bessel}
u_{(S)}^{far}=\sqrt{z}\bigg[\tilde{C}_{(S)}I_{\alpha+s}(z/2)+\tilde{D}_{(S)}I_{-\alpha-s}(z/2)\bigg]~,
\end{eqnarray}
where $\alpha=l+1/2$. One may then use the relations between adjacent Bessel functions \cite{Abramowitz} to show that
\begin{eqnarray} \label{even_solution_far_vector}
\psi^{far}&=&\sqrt{z}\bigg[\tilde{C}_VI_{\alpha}(z/2)+\tilde{D}_VI_{-\alpha}(z/2)\bigg]=\nonumber\\
&=&e^{-z/2}\bigg[C_Vz^{l+1}M(l+1,2l+2,z)+\nn\\ 
&+&D_Vz^{-l}M(-l,-2l,z)\bigg]~,
\end{eqnarray}
where, setting $c_{(2)}^\pm=1$ without loss of generality, we have
\begin{eqnarray} \label{even_solution_far_vector_constants}
\tilde{C}_V&=&-l\tilde{C}_{(+)} +(l+1)\tilde{C}_{(-)}~,\nn \\
\tilde{D}_V&=&-l\tilde{D}_{(+)} +(l+1)\tilde{D}_{(-)}~.
\end{eqnarray}
The corresponding constants $C_V$ and $D_V$ can be obtained from these using Eq.~(\ref{modified_bessel}). This shows that, within this approximation, $\psi$ has the same form as the odd-parity function $u_{(4)}$ both in the near-horizon and asymptotically flat regions and should hence yield the same QN mode spectrum. As our numerical analysis shows, this holds only in the massless limit, so that this approach fails to describe the broken degeneracy between the even- and odd-parity `vector' QN modes for finite $\mu$. It nevertheless illustrates how the two $j=l\pm1$ solutions combine to form `vector' states in the far-field region for small $\nu$.

One may also obtain the `scalar' QN modes corresponding to solutions with $\psi=0$. In the far region this implies from Eqs. (\ref{even_solution_far_vector}) and (\ref{even_solution_far_vector_constants}) that $\tilde{C}_{(+)}/\tilde{C}_{(-)}-=\tilde{D}_{(+)}/\tilde{D}_{(-)}=(l+1)/l$, and we may write
\begin{eqnarray} \label{even_solution_far_scalar}
u_{(3)}^{far}&=&2(l+1)(2l+1)z^{-1/2}\nn\\\ 
&\times&\bigg[\tilde{C}_{(-)}I_{l+1/2}(z/2)+\tilde{D}_{(-)}I_{-l-1/2}(z/2)\bigg]=\nonumber\\
&=&e^{-z/2}\bigg[C_{(3)}z^lM(l+1,2l+2,z)+\nn \\
&+&D_{(3)}z^{-l-1}M(-l-2l,z)\bigg]~.
\end{eqnarray}
We may then use the asymptotic properties of the modified Bessel functions \cite{Abramowitz} in this case to show that $u_{(3)}/u_{(2)}\rightarrow0$, in agreement with the numerical polarization data in the massless limit. Taking the inverse ratios $\tilde{C}_{(+)}/\tilde{C}_{(-)}=\tilde{D}_{(+)}/\tilde{D}_{(-)}=l/(l+1)$, we obtain $u_{(3)}/u_{(2)}\rightarrow1$ asymptotically, which correspond to the `vector' even-parity QN modes as obtained numerically.

We may derive a matching condition for the `scalar' QN modes in a similar fashion to the odd-parity case, by first matching the constants $C_{(3)}$ and $D_{(3)}$ to the corresponding coefficients in Eq.~(\ref{even_solution_near_large}) and then imposing outgoing waves at infinity. From Eq.~(\ref{even_solution_far_scalar}), we obtain asymptotically
\begin{eqnarray} \label{even_solution_far_scalar_large}
u_{(3)}^{far}&\simeq& {e^{-z/2}\over z}(-1)^{l+1}\times\nn \\ &\times&\!\!\!\!\bigg[C_{(3)}{\Gamma[2l+2]\over\Gamma[l+1]})-D_{(3)}{\Gamma[-2l]\over\Gamma[-l]}\bigg]+\nn \\
&+&{e^{z/2}\over z}\bigg[C_{(3)}{\Gamma[2l+2]\over\Gamma[l+1]}+D_{(3)}{\Gamma[-2l]\over\Gamma[-l]}\bigg]~.
\end{eqnarray}

Setting the coefficient of the first term to zero, we obtain the `scalar' matching condition
\begin{eqnarray} \label{even_QNM_matching_scalar}
1&=&(4q)^{2l+1}{\Gamma[-2l]\over\Gamma[-l]}{\Gamma[l+1]\over\Gamma[2l+2]}{\Gamma[-2l-1]\over\Gamma[2l+1]}{\Gamma[l+1]\over\Gamma[-l]}\times\nn \\ 
&\times&{\Gamma[l+1-4i\omega]\over\Gamma[-l-4i\omega]}~.
\end{eqnarray}
This can be further simplified to yield
\beq \label{even_QNM_matching_scalar_2}
4\omega(4\sqrt{\omega^2-\mu^2})^{2l+1}\!\!\left({(l!)^2\over (2l)!(2l+1)!}\right)^2\!\!\prod_{k=1}^l\!(k^2+16\omega^2)\!=\!-1~.
\eeq
For comparison with the `vector' QN modes, we solved this equation to leading order for the first excited state, yielding:
\begin{eqnarray} \label{even_QNM_scalar}
\omega_{(1,0)}^{even,S}\simeq 0.612(1-i)+0.306(1+i)\mu^2~,
\end{eqnarray}
which as before overestimates the numerical result but yields the correct qualitative $\mu$ dependence, also showing that the `scalar' QN modes have larger frequencies than the corresponding `vector' states, in agreement with our earlier results.

Hence, although the analytical matching procedure cannot really replace the numerical analysis in terms of quantitative results, it illustrates the rich and nontrivial structure of the massive vector field perturbations, in particular that of the even-parity modes. This interesting structure is a consequence of the different spin structure near and far from the black-hole horizon, leading to a nontrivial interplay between spin and orbital angular momentum. In particular, whereas for QN modes one finds `scalar' and `vector' states at infinity as in the massless case, these two components are nontrivially mixed for bound states, which have no massless counterpart.


\section{Conclusion and future prospects\label{sec:conclusion}}

As we have found, massive vector fields exhibit an extremely rich spectrum of perturbations on the Schwarzschild spacetime, due to both the nonvanishing mass and the spin-1 nature of the field. Our results show that the `vector' and `scalar' solutions describing electromagnetic perturbations on this geometry, the latter being unphysical gauge degrees of freedom in electromagnetism, mix in a nontrivial fashion in the presence of a nonzero mass, even for small-mass coupling $M\mu$. 

For the electromagnetic field ($\mu = 0$), (i) the even- and odd-parity `vector' modes are governed by the same dynamical equation, and hence their QN frequency spectra are degenerate, and (ii) the `scalar' degree of freedom corresponds to a `pure-gauge' mode. By contrast, in the Proca case ($\mu \neq 0$), as a consequence of the breaking of the electric-magnetic duality, we find that (i) the odd- and even-parity `vector' modes become dynamically distinct, and (ii) the even-parity `scalar' mode acquires a physical significance, since the Proca field has no gauge freedom ($A^{\mu}_{;\mu} = 0$). We showed in Sec.~\ref{subsec:separation} that the odd-parity part of the field is governed by a single equation, whereas the even-parity part of the field is determined by a coupled pair of equations. 

The Proca field exhibits quasibound states, i.e.~solutions which can be localized within the vicinity of the black-hole horizon and which are absent in the Maxwell case. Quasibound states on the Schwarzschild spacetime have complex frequencies, with an imaginary part corresponding to the decay rate (as flux is absorbed by the horizon). Our numerical and analytical studies of the bound-state spectra reveal an interplay between the `vector' and `scalar' solutions, which is reminiscent of a spin-orbit coupling between the field's proper spin and the orbital angular momentum of each multipole. States may be labeled by their total angular momentum $j=l+S$, as measured by an asymptotic observer. We thus find `electric' (even-parity) transverse states, with $j=l\pm1$, and `magnetic' (odd-parity) longitudinal states, with $j=l$, in both cases yielding a hydrogenic spectrum for $M\mu\ll1$ labeled by the `principal quantum number' $N=j+1+n$ for non-negative integers $n$. While this agrees with earlier studies for both the monopole \cite{K:2006} and large multipoles \cite{Galtsov}, we find  decay times which are parametrically different for each type of mode, as opposed to \cite{Galtsov} where a common behaviour for small-mass coupling $\text{Im}(\omega/\mu)\propto (M\mu)^{4l+5}$ was found for all spin-$j$ states, albeit with different coefficients. Although some further analytical insight is required to better understand this behaviour, from our numerical results we can infer a power-law behaviour $\text{Im}(\omega/\mu)\propto (M\mu)^{4l+2S+5}$ in the same limit.

The fact that the bound-state decay rate is sensitive to its spin in addition to its orbital angular momentum can be understood in simple physical terms in the `antitunneling' picture devised in \cite{Arvanitaki-JMR}. Bound states are localized in a potential well whose depth depends on the field's mass and total spin and which is separated from the black-hole horizon by a finite angular momentum barrier. The rate at which the black hole attenuates the mode's wavefunction is then determined by the height of this barrier, which is controlled by the {\it total} angular momentum of the state. In particular, the states with the smallest angular momentum $j=l-1$ for each multipole are absorbed more quickly by the black hole and hence exhibit a faster decay rate. For higher multipoles, we have nevertheless that $j\simeq l\gg1$, which justifies the results obtained in \cite{Galtsov}.

These results also suggest that a similar behaviour should be observed for rotating black holes, where wave modes with $\omega<m\Omega$ are amplified rather than damped by superradiant scattering in the Kerr ergoregion, with $-l\leq m\leq l$ denoting the azimuthal angular momentum projection and $\Omega$ the black hole's rotational frequency. When such states are bound to the black hole, multiple wave scatterings will amplify the corresponding wave function and consequently exponentially increase the associated particle number, giving rise to the so-called {\it black hole bomb} effect \cite{Press-1972}. The total angular momentum barrier should then also determine the overlap of each mode with the ergoregion and we thus expect the $j=l-1$ bound states of the Proca field to exhibit a parametrically faster instability rate in this case as well.

A rigorous analysis of the massive vector field bound states on the Kerr spacetime poses, however, an extremely challenging problem even from the numerical point of view as the field equations do not seem to admit separable solutions in this geometry. In addition, parity invariance is broken by the black hole's rotation, which may give rise to nontrivial mixings between the odd- and even-parity states, namely in the extremal case, where we expect the superradiant instability to be strongest \cite{Dolan-2007, Rosa}. One might expect these effects to become subdominant for slowly-rotating black holes, and a preliminary analysis in this limit seems to confirm our physical intuition, with the lowest spin states exhibiting a parametrically larger instability, although it remains unclear whether it is consistent to study superradiant modes in this case. 

Although a comprehensive study of Proca perturbations on the Kerr spacetime is beyond the scope of this work, the physical picture derived from our results for nonrotating geometries suggests that Kerr superradiance may be relevant for probing the existence of ultralight hidden $U(1)$ vector fields in string compactifications, as described in the introduction. In particular, one expects this effect to be more pronounced for hidden photons than for axionlike fields, given the existence of states with a lower angular momentum barrier due to the above mentioned spin-orbit coupling. In addition, the lowest-lying odd-parity longitudinal bound states of the Proca field exhibit a faster decay rate than the corresponding scalar-field modes in the Schwarzschild case as, despite yielding similar angular momentum barriers, higher-spin waves decay more rapidly in this case, a well-known result for massless fields \cite{Teukolsky:1974}. 

This suggests that, in the case where axionlike and hidden photons of similar masses coexist, superradiant scattering will amplify the latter bound states more quickly. This will in turn inhibit the formation of axionlike bound states, given that the hidden photon cloud breaks the rotational symmetry of the system and thus suppresses multiple scatterings in states with distinct quantum numbers, as described in \cite{Arvanitaki}. The cloud may later be depleted by self-interactions and the progressive shutdown of the mode instability due to the decreasing black hole mass, so that other states may be amplified, possibly leading to an interplay between spin-1 and spin-0 states with interesting phenomenological consequences. Furthermore, similarly to the axion case, massive hidden vector field clouds around astrophysical black holes should also lead to phenomena such as gravitational waves and possibly `bosenova-like' emission, while the generic mixing between hidden and visible photons could yield exciting novel signatures \cite{Arvanitaki-JMR, Arvanitaki, Kodama:Yoshino:2011}.

The study of massive vector field perturbations in black hole spacetimes is thus an important problem from both the black hole stability and phenomenological perspectives. We hope in the future to further develop and extend our analytical and numerical methods to better understand the behaviour of massive higher-spin perturbations on the Schwarzschild spacetime, as well as for more generic black hole geometries with angular momentum and charge.


\acknowledgements

We thank John March-Russell and Leor Barack for useful discussions on this topic. J.G.R. is supported by STFC. S.R.D. acknowledges support from EPSRC through Grant No. EP/G049092/1.




\begin{thebibliography}{20}


\bibitem{Regge-Wheeler:1957}
T. Regge and J. A. Wheeler, 
Phys. Rev. {\bf 108}, 1063 (1957).

\bibitem{Vishveshwara:1970} 
C. V. Vishveshwara, Nature (London) 227, 936 (1970).

\bibitem{Bekenstein:1972}
J.~D.~Bekenstein, Phys. Rev. D {\bf 5}, 1239 (1972).

\bibitem{Page:1976}
D. N. Page,
Phys.~Rev.~D {\bf 13}, 198 (1976).

\bibitem{Matzner:1968}
R. A. Matzner, J. Math Phys. {\bf 9}, 163 (1968).

\bibitem{Dolan:2008}
S.~R.~Dolan, Class. Quantum Grav. {\bf 25}, 235002 (2008) [arXiv:0801.3805]. 

\bibitem{Crispino:Dolan:Oliveira:2009}
L.~C.~B.~Crispino, S.~R.~Dolan and E.~S.~Oliveira, Phys.~Rev.~Lett. {\bf 102}, 231103 (2009) [arXiv:0905.3339].

\bibitem{Zerilli:1970}
F. J. Zerilli, 
Phys. Rev. D {\bf 2}, 2141 (1970).

\bibitem{Chandrasekhar:Detweiler:1975}
S. Chandrasekhar and S. Detweiler,
Proc. Roy. Soc. Lond. A {\bf 344}, 441 (1975).

\bibitem{Berti-Cardoso-Starinets}
E.~Berti, V.~Cardoso and A.~O.~Starinets,
Class.~Quantum Grav. {\bf 26} (2009) 163001 [arXiv:0905.2975].

\bibitem{Konoplya:2011}
  R.~A.~Konoplya, A.~Zhidenko,
  Rev.\ Mod.\ Phys.\  {\bf 83}, 793-836 (2011)
  [arXiv:1102.4014 [gr-qc]].

\bibitem{Price:1972}
R. H. Price, Phys. Rev. D {\bf 5}, 2439 (1972).

\bibitem{Deruelle-1974}
N. Deruelle and R.~Ruffini,
Phys. Lett. B {\bf 52} (1974), 437--441.

\bibitem{Damour-1976}
T. Damour, N. Deruelle and R. Ruffini,
Lett. Nuovo Cimento {\bf 15} (1976), 257.

\bibitem{Chandrasekhar:1976}
S. Chandrasekhar
Proc. R. Soc. Lond. A {\bf 349}, 571 (1976).

\bibitem{Gal'tsov-1983}
D.~V.~Gal'tsov, G.~V.~Pomerantseva, and G.~A.~Chizhov,
Sov. Phys. J. {\bf 26}, 743 (1983).

\bibitem{Galtsov}
D. V. Gal'tsov, G. V. Pomerantseva and G. A. Chizhov, Sov. Phys. J. {\bf 27}, 697 (1984) [Izv. Vuz. Fiz. {\bf 27}, 81 (1984)].

\bibitem{Gaina-1992}
A.~B.~Gaina and O.~B.~Zaslavskii,
Class. Quantum Grav. {\bf 9}, 667 (1992).

\bibitem{Simone-1992}
L.~E.~Simone and C.~M.~Will,
Class. Quantum Grav. {\bf 9}, 963 (1992).

\bibitem{Gaina-1993}
A.~B.~Gaina, and N.~I.~Ionescu-Pallas, 
Rom. J. Phys. {\bf 38}, 729 (1993).

\bibitem{Lasenby-2005-bs}
A. N. Lasenby, C. J. L. Doran, J. Pritchard, A.~Caceres and S.~R.~Dolan,
Phys. Rev. D {\bf 72}, 105014 (2005).

\bibitem{Doran-2005}
C. J. L. Doran, A. N. Lasenby, S. R. Dolan and I. Hinder,
Phys. Rev. D {\bf 71}, 124020 (2005).

\bibitem{Dolan-2006}
S. R. Dolan, C. J. L. Doran, A. N. Lasenby,
Phys. Rev. D {\bf 74}, 064005 (2006).

\bibitem{Laptev-2006}
Y.~P.~Laptev and M.~L.~Fil'chenkov,
Astronomical and Astrophysical Transactions, {\bf 25}, 33 (2006).

\bibitem{Grain-Barrau}
J.~Grain and A.~Barrau,
Eur.~Phys.~J.~C {\bf 53}, 641 (2008) [arXiv:hep-th/0701265].

\bibitem{KerrFermi}
T.~Hartman, W.~Song, A.~Strominger, (2009)
[arXiv:0912.4265].

\bibitem{Koyama:2001}
H. Koyama and A. Tomimatsu, 
Phys.~Rev.~D {\bf 64}, 044014 (2001).

\bibitem{Koyama:2002}
H. Koyama and A. Tomimatsu, 
Phys.~Rev.~D {\bf 65}, 084031 (2002).

\bibitem{Jing:2005}
J. Jing, 
Phys. Rev. D {\bf 72}, 027501 (2005).

\bibitem{KZ:2006}
  R.~A.~Konoplya, A.~Zhidenko,
  Phys.\ Rev.\  {\bf D73}, 124040 (2006).
  [gr-qc/0605013].

\bibitem{K:2006}
R.~A.~Konoplya, 
Phys.~Rev.~D {\bf 73}, 024009 (2006).

\bibitem{KZM:2007}
R.~A.~Konoplya, A.~Zhidenko and C.~Molina, 
Phys.~Rev.~D {\bf 75}, 084004 (2007).

\bibitem{Hawking}
S.~Hawking,
Mon.~Not.~R.~Astron.~Soc. {\bf 152}, 75 (1971).

\bibitem{Zeldovich}
Y.~B.~Zel’dovich and I.~D.~Novikov,
Astron.~Zh.~{\bf 43}, 758 (1966); Sov.~Astron.~{\bf 10}, 602 (1967).

\bibitem{Carr}
B.~J.~Carr and S.~W.~Hawking,
Mon.~Not.~R.~Astron.~Soc. {\bf 168}, 399 (1974).

\bibitem{Press-1972}
W.~H.~Press and S.~A.~Teukolsky,
\newblock {\em Nature \bf{238}\/} (1972), 211.

\bibitem{Teukolsky:1973}
S.~A.~Teukolsky, Astrophys.~J.~{\bf 185}, 635 (1973).

\bibitem{Teukolsky:1974}
S.~A.~Teukolsky and W.~H.~Press,
Astrophys.\ J.\  {\bf 193}, 443-461 (1974).

\bibitem{Zouros:1979}
T.~Zouros and D.~Eardley,
Annals of Physics {\bf 118}, 139 (1979).

\bibitem{Detweiler-1980}
S.~Detweiler,
Phys. Rev. D {\bf 22}, 2323 (1980).

\bibitem{Cardoso:2004}
V.~Cardoso, O.~J.~C. Dias, J.~P.~S. Lemos and S.~Yoshida
Phys. Rev. D {\bf 70}, 044039 (2004).

\bibitem{Furuhashi:2004jk}
H.~Furuhashi and Y.~Nambu,
Prog.\ Theor.\ Phys.\  {\bf 112}, 983 (2004)
[arXiv:gr-qc/0402037].

\bibitem{Cardoso:2005vk}
  V.~Cardoso, S.~Yoshida,
  JHEP {\bf 0507}, 009 (2005)
  [hep-th/0502206].

\bibitem{Dolan-2007}
S.~R.~Dolan, 
Phys. Rev. D {\bf 76}, 084001 (2007).
.

\bibitem{Konoplya:2008}
R. A. Konoplya,
Phys. Lett. B {\bf 670}, 459 (2009).

\bibitem{Rosa}
J.~G.~Rosa,
JHEP {\bf 06}, 015 (2010) [arXiv:0912.1780].

\bibitem{Cardoso:2011xi}
  V.~Cardoso {\it et al.}, 
  Phys.\ Rev.\ Lett.\  {\bf 107}, 241101 (2011)
  [arXiv:1109.6021 [gr-qc]].

\bibitem{Arvanitaki-JMR}
A.~Arvanitaki {\it et al.}, 
Phys. Rev. D {\bf 81}, 123530 (2010) [arXiv:0905.4720].

\bibitem{Arvanitaki}
A. Arvanitaki and S. Dubovsky,
Phys.~Rev.~D {\bf 83}, 044026 (2011) [arXiv:1004.3558].

\bibitem{Kodama:Yoshino:2011}
H. Kodama and H. Yoshino, (2011)
arXiv:1108.1365.

\bibitem{Lora-2011}
V. Lora, J. Magana, A. Bernal, F. J. Sanchez-Salcedo and E. K. Grebel (2011), arXiv:1110.2684.

\bibitem{Goodsell}
M.~Goodsell, J.~Jaeckel, J.~Redondo, A.~Ringwald,
JHEP {\bf 0911}, 027 (2009)
[arXiv:0909.0515].

\bibitem{Jaeckel}
J.~Jaeckel, A.~Ringwald,
Ann.\ Rev.\ Nucl.\ Part.\ Sci.\  {\bf 60 } (2010)  405-437
[arXiv:1002.0329].

\bibitem{Camara}
P.~G.~Camara, L.~E.~Ibanez, F.~Marchesano,
JHEP {\bf 1109}, 110 (2011)
[arXiv:1106.0060 [hep-th]].


\bibitem{ArkaniHamed}
N.~Arkani-Hamed, L.~Motl, A.~Nicolis, C.~Vafa,
JHEP {\bf 0706}, 060 (2007)
[hep-th/0601001].

\bibitem{PDG}
K.~Nakamura {\it et al.} [Particle Data Group Collaboration],
J.\ Phys.\ G {\bf 37}, 075021 (2010).

\bibitem{Tamburini}
F.~Tamburini, A.~Sponselli, B. Thid�e and J. T. Mendonca,
EPL {\bf 90}, 45001 (2010).

\bibitem{Anderson}
P.~W.~Anderson, Phys. Rev. {\bf 130}, 439 (1963).

\bibitem{Herdeiro}
C. Herdeiro, M. O. P. Sampaio and M. Wang,
Phys.\ Rev.\ D {\bf 85}, 024005 (2012)
[arXiv:1110.2485 [gr-qc]].





\bibitem{Barack:Lousto:2005}
L.~Barack and C.~O.~Lousto,
Phys. Rev. D {\bf 72}, 104026 (2005). 

\bibitem{Chandrasekhar:1975}
S.~Chandrasekhar, 
Proc. Roy. Soc. Lond. A {\bf 343}, 289 (1975).

\bibitem{Chandrasekhar:1983}
S. Chandrasekhar,
\emph{The Mathematical Theory of Black Holes} (New York: Oxford University Press, 1992).

\bibitem{Kinnersley}
  W.~Kinnersley,
  J.\ Math.\ Phys.\  {\bf 10}, 1195-1203 (1969).



\bibitem{Leaver-1985}
E.~W.~Leaver,
Proc. R. Soc. London A {\bf 402}, 285 (1985).

\bibitem{Zhidenko-thesis}
A.~Zhidenko,
Ph.D thesis, IFUSP, Sao Paulo (2009) [arXiv:0903.3555].


\bibitem{SWP:1999}
C.~Simmendinger, A.~Wunderlin, A.~Pelster, 
Phys. Rev. E {\bf 59}, 5344 (1999).

\bibitem{Abramowitz}
M. Abramowitz and I. A. Stegun, eds., {\it Handbook of Mathematical
Functions With Formulas, Graphs, and Mathematical Tables}, NBS App.~Math.~Series {\bf 55}, National Bureau of
Standards, Washington, DC (1964).

\bibitem{Panchapakesan:1987}
N.~Panchapakesan and B.~Majumdar,
Astrophys.\ Space\ Sci.\ {\bf 136}, 251 (1987).





\end{thebibliography}

\end{document}